\documentclass[aps, prb, amsmath,amssymb,
 preprint,
 superscriptaddress,
]{revtex4-2}

\usepackage{graphicx}
\usepackage{dcolumn}
\usepackage{bm}
\usepackage[mathlines]{lineno}
\usepackage{color}
\usepackage{longtable}
\usepackage[colorlinks=true,allcolors=blue]{hyperref}
\usepackage{ulem}

\begin{document}
\newcommand{\micron}{$\mu$m}

\title{Toward an accurate equation of state and B1-B2 phase boundary for magnesium oxide to TPa pressures and eV temperatures}

\author{Shuai Zhang}
\email{szha@lle.rochester.edu}
\affiliation{Laboratory for Laser Energetics, University of Rochester, Rochester, New York 14623, USA}

\author{Reetam Paul}
\affiliation{Laboratory for Laser Energetics, University of Rochester, Rochester, New York 14623, USA}
\affiliation{Lawrence Livermore National Laboratory, Livermore, California 94550, USA}


\author{S. X. Hu}
\affiliation{Laboratory for Laser Energetics, University of Rochester, Rochester, New York 14623, USA}

\author{Miguel A. Morales}
\email{mmorales@flatironinstitute.org}
\affiliation{Lawrence Livermore National Laboratory, Livermore, California 94550, USA}
\affiliation{Center for Computational Quantum Physics, Flatiron Institute, New York, New York 10010, USA}

\date{\today}

\begin{abstract}
By applying auxiliary-field quantum Monte Carlo, we calculate the equation of state (EOS) and B1-B2 phase transition of magnesium oxide (MgO) up to 1 TPa. 
The results agree with available experimental data at low pressures and are used to benchmark the performance of various exchange-correlation functionals in density functional theory calculations.
We determine PBEsol is an optimal choice for the exchange-correlation functional and perform extensive phonon and quantum molecular-dynamics calculations to obtain the thermal EOS.
Our results provide a preliminary reference for the EOS and B1-B2 phase boundary of MgO from zero up to 10,500~K.
\end{abstract}

\maketitle

\section{Introduction}
Materials structures and behaviors under very high pressure ($\sim$100 GPa to 1 TPa) are an important topic in high-energy-density sciences and earth and planetary sciences.
At such conditions, materials are strongly compressed, which can lead to transitions into phases with different structures (by lowering the thermodynamic energies) and chemistry (by modifying the bonding). 
The past two decades have seen advances in computing and compression technologies that have added important knowledge to this subject by unveiling new structures (e.g., MgSiO$_3$ post-perovskite~\cite{Murakami2004,Oganov2004,Tsuchiya2004})
or chemical stoichiometry (such as H$_4$O~\cite{Zhang2013h4o}
and Xe-FeO$_2$~\cite{Peng2020}) with notable changes to properties of chemical systems (particularly the insulator-metal transition~\cite{Celliers2018,Rillo2019} and high-temperature superconductivity~\cite{Zurek2019}).
However, accurate determination of phase transitions at such extreme conditions remains challenging.

Experimentally, static compression experiments based on diamond-anvil cells (DACs)~\cite{Dubrovinskaia2016,Jenei2018} are limited by sample sizes and diagnostics, while dynamic compression experiments are limited by the time scale and regime of the thermodynamic paths that can be achieved~\cite{McWilliamsScience2012,Coppari2013,MillotSci2015}.

Theoretically, state-of-art investigations often rely on calculations based on Kohn--Sham density functional theory (DFT)~\cite{dft2ks}.
Despite the tremendous success of DFT in predicting many structures and properties to moderately high pressures, errors associated with the single-particle approximation and exchange-correlation (xc) functionals render DFT predictions dubious where precise experimental constraints do not exist.

Recent studies have shown quantum Monte Carlo (QMC) methods
to be able to benchmark solid-state equation of state (EOS)
and phase transitions~\cite{Zhang2018NiO,Driver2010SiO2,Trail2017TiO2}
by directly solving the many-electron Schr\"{o}dinger
equation.
Auxiliary-field quantum Monte Carlo (AFQMC) is one
such QMC method that has shown great promise 
with flexibility and scalability for simulating both real and model many-body systems with high accuracy~\cite{Zhang2018NiO,Malone2019C,Malone2020,PhysRevX.5.041041,PhysRevX.7.031059,Zheng1155,Lee2020jcp,Shi2021jcp}. 


In this work we apply the phaseless AFQMC~\cite{Zhang2003phaseless,motta2017ab} method, in combination with optimized periodic Gaussian basis sets~\cite{Zhang2018NiO,Malone2019C,Morales2020}, to investigate high-pressure EOS and phase transition in solid-state materials by using magnesium oxide (MgO) as an example.
This provides theoretically accurate cold-curve results for MgO, which we then use to benchmark against various predictions by DFT calculations.
We then use DFT-based lattice dynamics and molecular dynamics with one of the best xc functionals to calculate the thermal contributions to the EOS.
Finally, we combined the thermal with the cold-curve results to determine the finite-temperature EOS and B1-B2 phase boundary for MgO to eV temperatures.

MgO is a prototype rock-forming mineral in planets, a pressure calibrator in DAC experiments, and a window material in shock experiments. 
From ambient pressure up to about 500 GPa, MgO is stabilized in the sodium chloride (NaCl, or B1) structure. Beyond that, it transforms into the cesium chloride (CsCl, or B2) structure, which is characterized by smaller coordination and lower viscosity that may be associated with mantle convection and layering in super-Earths different from those in the Earth.
A benchmark of the EOS and phase transition of MgO would be important for modeling the interior dynamics and evolution of super-Earths,	testing the degree of validity of various theoretical EOS or models at extreme conditions, as well as elucidating materials physics at extreme conditions by answering such questions as: is thermodynamic equilibrium reached in the experiments, or to what degree are the phase transformations subject to kinetic effects, chemistry/composition changes, or a combination of them, leading to various observations in experiments?
This problem of the B1-B2 transition in MgO has been studied for over 40 years but remains uncertain in experiments~\cite{Coppari2013,Dubrovinskaia2009} and there is a discrepancy of $\sim$20\% between state-of-the-art DFT calculations~\cite{Bouchet2019}.
In addition to the debates over phase relations near the triple point near 500 GPa~\cite{Alfe2005mgomelt,dKS2009,Belonoshko2010,Boates2013,Cebulla2014,Taniuchi2018,Musella2019,Bouchet2019,Soubiran2020,McWilliamsScience2012,Root2015mgo,Miyanishi2015,Bolis2016mgo}, 
recent double-shock experiments also suggest an inconsistency exists between theoretical predictions and experiments of the melting curve at TPa pressures~\cite{Hansen2021}.

The main goal of this work is to provide an accurate EOS and phase diagram for MgO to TPa pressures and eV temperatures by
jointly combining an accurate many-body electronic structure approach (AFQMC)
and finite-temperature quantum molecular dynamics (QMD) based on DFT
to 
fully address various details of physics (electronic correlation, anharmonic vibration, EOS models, finite sizes of the simulation cell, and Born effective charge) that can affect the thermal EOS results.



This paper is organized as follows:
Section~\ref{sec:method} outlines the methodologies used in this study, including 
those for zero-K and finite-temperature calculations;
Sec.~\ref{sec:result} presents the cold curve, thermal EOS, and phase boundary results for MgO, and discusses the errors and their sources; 
finally, Sec.~\ref{sec:conclusions} concludes the paper.

\section{Methods}\label{sec:method}
In the following, we present descriptions and settings of the computational approaches used in this study, including AFQMC, Hartree--Fock (HF), and DFT for the zero-K internal-energy calculations, and quantum molecular dynamics (QMD) and thermodynamic integration (TDI) for the thermodynamic free energies at nonzero temperatures.

\subsection{Zero-K static lattice calculations}\label{subsec2:coldcurve}
For the internal energy-volume $E(V)$ relations at 0 K (often called the ``cold curve''), we perform static lattice calculations for MgO in the B1 and B2 structures at a series of volumes by using a combination of AFQMC, HF, and DFT with various xc functionals.


AFQMC is a zero-temperature quantum Monte Carlo approach. It is based on the stochastic propagation of wavefunctions in imaginary time using an ensemble of walkers in the space of non-orthogonal Slater determinants.
It uses the Hubbard--Stratonovich transformation~\cite{PhysRevLett.3.77} to rewrite the expensive two-body part of the propagator into an integral over auxiliary fields coupled to one-body propagators, which are then integrated with Monte Carlo techniques.
Like other QMC methods, AFQMC also faces an obstacle for fermionic systems, namely the ``phase (or sign) problem'' which arises because the fields led by Coulomb interaction are complex. Control of the sign problem can be achieved using constraints based on trial wavefunctions, like the fixed-node approximation in diffusion MC (DMC)~\cite{Ceperley1980} or, in the case of AFQMC, the constrained-path~\cite{zhang_cpmc} and phaseless approximation~\cite{Zhang2003phaseless}. When combined with appropriate trial wavefunctions, these methods have been shown to provide benchmark-quality results across a range of electronic structure problems including atoms, molecules, solids, and correlated model Hamiltonians, including cases with strong electronic correlations~\cite{PhysRevX.5.041041,Zheng1155,Shee2019}, known to be challenging to alternative approaches like Kohn--Sham DFT.
Recent advances in the development of accurate and flexible trial wavefunctions include the use of multi-determinant expansions~\cite{doi:10.1063/1.3077920,borda2018non,Mahajan2021} and generalized HF~\cite{PhysRevB.94.085103,chang2017multi}.
In this work, we use the phaseless AFQMC (ph-AFQMC) method~\cite{Zhang2003phaseless,motta2017ab} to calculate the ground state properties of bulk MgO.

In our {\color{black}ph-AFQMC} calculations, the trial wavefunction is constructed from a Slater determinant of HF orbitals (the HF solution for each MgO structure at every density), which were found to yield accurate energy results.
We use {\footnotesize QUANTUM ESPRESSO} {\color{black}(QE)}~\cite{QE2009,QE2017} for the calculation of the trial wavefunction and for the generation of the one- and two-electron integrals.
The modified Cholesky decomposition~\cite{beebe77,doi:10.1063/1.1578621,doi:10.1002/jcc.21318,doi:10.1063/1.3654002} is used to avoid the $\mathcal{O}(M^4)$ cost of storing the two-electron repulsion integrals.
All {\color{black}QE} simulations were performed using optimized norm-conserving Vanderbilt (ONCV) pseudopotentials~\cite{PhysRevB.88.085117}, constructed with the Perdew--Burke--Ernzerhof (PBE)~\cite{PBE1996} xc functional.
We used the recently developed optimized Gaussian basis sets~\cite{Morales2020} in all AFQMC calculations.
{\color{black}The calculations were based on primitive unit cells and performed using $\Gamma$-centered 2$\times$2$\times$2, 3$\times$3$\times$3, and 4$\times$4$\times$4 $k$ grids to extrapolate to the thermodynamic limit at each density.
Results from multiple basis sets were used,
in combination with corrections based on periodic second-order M\o ller--Plesset perturbation theory (MP2) calculations, to obtain results extrapolated to the complete basis set (CBS) limit (see Appendix~\ref{secapp:afqmcconverge} for more details)}.
This was shown to be a successful approach to removing basis and finite size errors in previous studies, to which we refer readers for additional details~\cite{Morales2020}.
All AFQMC calculations were performed using the open-source {\footnotesize QMCPACK} software package~\cite{qmcpack}.
We used $\sim$1000 walkers and a time step of 0.005~{Ha}$^{-1}$, which we found sufficient to control any potential population and finite time-step biases, respectively.

Kohn--Sham DFT~\cite{dft2ks} follows the Hohenberg--Kohn theorem~\cite{dft1hk} and simplifies the many-body problem into a single-particle mean-field equation that can be solved self-consistently via iteration over the electron density.
The real complicated electron-electron interactions are simplified into the xc functional term.
Since the accurate QMC solution for the uniform electron gas~\cite{Ceperley1980},
there have been developments of many forms of xc functionals for various applications,
which form a ``Jacob's ladder'' with different rungs [local density approximation (LDA), generalized gradient approximation (GGA), meta-GGA, etc.] that lead to chemical accuracy at the expense of increasing computational cost.

{\color{black}Our} DFT calculations {\color{black}of the cold curve} are performed with Vienna \textit{Ab initio} Simulation Package ({\footnotesize VASP})~\cite{kresse96b}.
In our {\footnotesize VASP} simulations, we use a two-atom unit cell,
    $\Gamma$-centered 16$\times$16$\times$16 Monkhorst--Pack $k$ mesh, a plane-wave
    basis with cutoff of 1200 eV, and convergence criteria of $10^{-7}$
    eV/cell for the self-consistent iteration. The simulations use the projector augmented wave (PAW)~\cite{Blochl1994} method with pseudopotentials labeled with ``sv\_GW'' and ``h'', 1.75 and 1.1-Bohr core radii, and treating the outermost 10 and 6 electrons as valence for Mg and O, respectively. 
    We consider five different xc functionals: LDA~\cite{LDApz1981}, PBE~\cite{PBE1996}, PBEsol~\cite{PBEsol}, strongly constrained and appropriately normed meta-GGA (SCAN)~\cite{SCAN}, and the Heyd--Scuseria--Ernzerhof-type HF/DFT hybrid functional (HSE06)~\cite{HSE2006}.



The DFT calculations also produce pressures that are not directly available from our AFQMC calculations because of the difficulties in QMC to calculate forces.
For consistency in data comparison between different approaches and determination of the B1-B2 transition, we fitted the $E(V)$ data to EOS models that are widely used in high-pressure studies.
{\color{black}It has long been known that high-order elastic moduli may be required to parameterize materials EOS under extreme (e.g., near 2-fold) compression~\cite{Jeanloz1988EOS}.}
Therefore, we have considered multiple EOS models and cross-checked them with a numerical (spline fitting) approach to ensure the accuracy of the EOS and phase-transition results.

We have considered two different analytical EOS models: 
one is the Vinet model~\cite{vinet}, which follows
\begin{equation}
E(V)=E(V_0)+\int_V^{V_0}P(V)\mathrm{d}V,
\label{eq:vinet}
\end{equation}
with \begin{equation}
P(V)=3B_0\frac{1-x}{x^2}e^{1.5(B_0'-1)(1-x)},
\label{eq:vinetP}
\end{equation}
where $x=(\frac{V}{V_{0}})^{1/3}$ 
and $V_0$ and $B_0'$ are, respectively, the volume and first-order pressure derivative of the bulk modulus at zero pressure;
the other is the Birch--Murnaghan model~\cite{Birchbm4} to the fourth order, which follows
\begin{equation}
E(V) = E_{0} + 9B_0V_0(f^2/2 + a_1f^3/3 + a_2 f^4/4),
\label{eq:bm4}
\end{equation} 
where $f = [(\frac{V_{0}}{V})^{2/3}-1]/2$ is the Eulerian finite strain, $a_1 = 1.5(B_0' - 4)$, and $a_2$ is another parameter.
We have also tested the third-order Birch--Murnaghan model, which does not include the $a_2$ term in Eq.~\ref{eq:bm4}, for comparison with the other models in selected cases (see Appendix~\ref{secapp:eosfit}).

\subsection{Finite-temperature thermodynamic calculations}\label{subsec2:highT}
Thermodynamic calculations at nonzero temperatures are performed in two different ways: one is from lattice dynamics by using the quasiharmonic approximation (QHA), and the other is based on QMD.

Within QHA, lattice vibrations are considered to be dependent on volume but independent of temperature.
In practice, one can use the small-displacement approach or density functional perturbation theory (DFPT) to calculate phonons at 0 K and then compute the thermodynamic energies analytically from quantum statistics.
Despite its wide usage and success in giving improved thermodynamic properties over the fully harmonic approximation for materials at relatively low temperatures, the applicability of QHA is questionable at high temperatures and for systems with light elements at low temperatures.
In comparison, QMD simulations significantly improve the description of lattice dynamics by naturally including all anharmonic vibrations.
By employing a TDI approach, the free energies can also be accurately calculated, which makes it possible to chart the phase-transition boundaries at finite temperatures.

We use the {\footnotesize PHONOPY} program~\cite{phonopy,phonopy-phono3py-JPSJ} and {\footnotesize VASP} to calculate the phonons of MgO at 0 K with DFPT and under QHA.
We have tested the effects of including the Born effective charge (which is necessary to correctly account for the splitting between longitudinal and transverse optical modes) and different xc functionals on the phonon band structures and vibrational energies (see Appendices~\ref{secapp:loto} and~\ref{secapp:xc}).
The calculation is performed at a series of volumes $V$.
This allows estimation of the ion thermal contributions $F_{\mathrm{i}-\mathrm{th}}(V, T)$, at any temperature $T$, to be added to the free energies $F(V,T)$ via
\begin{align} 
F_{\mathrm{QHA}}(V,T)
& = E_{\mathrm{QHA}}(V,T)-T S_{\mathrm{QHA}}(V,T) \nonumber \\
& = k_{\mathrm{B}} T \sum_{\boldsymbol{q}, s} \ln \left[2 \sinh \left(\hbar \omega_{\boldsymbol{q}, s} / 2 k_{\mathrm{B}} T\right)\right],
\end{align}
where
$E_{\mathrm{QHA}}(V,T)
= \sum_{\boldsymbol{q}, s} \left( \tilde{n}+1/2\right) \hbar\omega_{\boldsymbol{q}, s}$ is the vibrational internal energy,
$\tilde{n}  =1/\left({e^{\hbar \omega_{\boldsymbol{q}, s} / k_{\mathrm{B}} T}-1}\right)$ is the effective number of the phonon mode with frequency $\boldsymbol{q}$ and index $s$,
and 
$S_{\mathrm{QHA}} (V,T)= k_{\mathrm{B}} \sum_{\boldsymbol{q}, s} \left[ \left( \tilde{n}+1\right) \ln \left( \tilde{n}+1\right) - \tilde{n}\ln\tilde{n}\right] $ is the vibrational entropy.
Each calculation employed a 54-atom supercell and was performed using a $\Gamma$-centered 4$\times$4$\times$4 $k$ mesh (for both B1 and B2 phases).

In QMD calculations, we use the Mermin--Kohn--Sham DFT approach~\cite{mermin1965} with the PBEsol xc functional. Ion temperatures are controlled by using the Nos\'e--Hoover thermostat~\cite{Nose1984}, while electron temperatures are defined by the Fermi--Dirac distribution via a smearing approach.
$NVT$ ensembles are generated that consist of 4000 to 10,000 MD steps with time step of 0.5 fs.
Mg\_sv\_GW and O\_h potentials are used, the same as the DFT calculations at 0 K.
The energy cutoff is 1000 eV, which defines the size of the plane-wave basis set.
It requires large enough cells in combination with proper/fine $k$ meshes to ensure the accuracy of the DFT calculations (see Appendix~\ref{secapp:convergence}).
In our simulations, we use cubic cells with 64 and 54 atoms that are sampled by a special $k$ point $(1/4,1/4,1/4)$ and $\Gamma$-centered 2$\times$2$\times$2 $k$ mesh for B1 and B2 phases, respectively{\color{black},
in order to obtain results reasonably close to the converged setting while computational cost is relatively low.}
Structure snapshots have been uniformly sampled from each QMD trajectory and recalculated with denser $k$ meshes of 2$\times$2$\times$2 (for B1) and 3$\times$3$\times$3 (for B2) to improve the accuracy of the thermal EOS and their volume dependence and reduce the error in the calculation of the phase transition.
The QMD calculations are performed at temperatures between 500 and 12,000 K, in steps of 500 to 1500 K, with more calculations at low to intermediate temperatures to improve the robustness of the TDI for anharmonic free-energy calculations.
Large numbers of 360 and 320 electronic bands are considered, respectively, for B1 and B2 simulations to ensure the highest-energy states remain unoccupied.
The EOS obtained from the QMD or QHA calculations produces $E(V,T)$ and $P(V,T)$ data that allow the calculation of the Hugoniot.
The analysis of the QMD EOS data follows the procedure that was introduced in detail in our recent paper on liquid SiO$_2$~\cite{Zhang2022sio2}.
The Hugoniot is calculated by solving the Rankine--Hugoniot equation using the numerically interpolated EOS.
The different theoretical predictions are then compared to the experimentally measured Hugoniot to benchmark the performance of the computational approaches and the xc functionals in the corresponding thermodynamic regime.




With the assistance of QHA and QMD, the entire ion thermal contribution to the free energy can be calculated by
\begin{equation} 
F_{\mathrm{i}-\mathrm{th}}(V, T)=F_{\mathrm{QHA}}(V, T)+F_{\mathrm{anharm}}(V, T).\label{eq:Fith}
\end{equation}
In Eq.~\ref{eq:Fith},
\begin{equation} 
F_{\mathrm{anharm}}(V, T)=-T \int_{T_{\mathrm{ref}}}^{T} \frac{E_{\mathrm{anharm}}(V, \mathcal{T}) 
}{\mathcal{T}^{2}} \mathrm{d} \mathcal{T}\label{eq:tdi}
\end{equation}
denotes the anharmonic term as calculated by TDI, where 
$E_{\mathrm{anharm}} = E_{\mathrm{QMD}}-E_{\mathrm{cold+QHA}}$, $E_{\mathrm{QMD}}$ is the internal energy from QMD simulations, and
$T_\mathrm{ref}$ is a reference temperature.

We note that QMD misses the quantum zero-point motion of ions while QHA does not. This leads to increased discrepancy between QMD and QHA internal energies as temperature drops near zero, associated with decreasing heat capacity $C_\mathrm{V}$ of the real system (from $\sim3k_\mathrm{B}$/atom to zero, as captured by QHA, whereas QMD gives $C_\mathrm{V}=3k_\text{B}$) since fewer lattice vibration modes can be excited.
In order to eliminate the resultant artificial exaggeration of the integrand, we have replaced $E_{\mathrm{cold+QHA}}$ with $E_{\mathrm{cold}}+3k_\mathrm{B}T$ in our calculations of Eq.~\ref{eq:tdi} (see Appendix~\ref{secapp:Fah}).
This effectively treats the ions classically in the evaluation of $F_{\mathrm{anharm}}$ at temperatures higher than $T_\text{ref}$, which we believe is a reasonable approximation for phonon interactions (the anharmonic term).

Our calculated results for $E_{\mathrm{anharm}}$
as a function of temperature are then fitted to high-order polynomials (to the sixth-order for B1 and eighth-order for B2) 
to compute the numerical integration in Eq.~\ref{eq:tdi}.
The functionality of TDI also requires choosing the proper reference point $T_\mathrm{ref}$. 
In this work, we consider $T_\mathrm{ref}$ to be low by following the idea that QHA is valid and other anharmonic contributions (beyond the volume-dependence vibration changes as have been included in QHA) are zero for MgO at low temperatures.
For consistency among different isochores, we make the choice of $T_\mathrm{ref}$ such that the heat capacity is 10\% of 3$k_\text{B}$. The corresponding $T_\mathrm{ref}$ is 100 to 200~K. We have also tested other choices of $T_\mathrm{ref}$ and examined their effects on $F_{\mathrm{anharm}}$ and the B1-B2 phase boundary. The results are summarized in Appendix~\ref{secapp:Fah}.

We note that when analyzing the QMD trajectory to calculate the EOS, we disregarded the beginning part (20\%) of each MD trajectory to ensure the reported EOS represents that under thermodynamic equilibrium.
Ion kinetic contributions to the EOS are manually included by following an ideal gas formula (i.e., internal energy $E_\textrm{ion kin.}=3Nk_\textrm{B}T/2$ and pressure $P_\textrm{ion kin.}=Nk_\textrm{B}T/V$, where $N$ is the total number of atoms in the simulation cell and $k_\text{B}$ is the Boltzmann constant).
Although MgO is an insulating solid with a wide electronic band gap in all conditions considered in this study, we have still carefully considered the effect of electron thermal effects in the free-energy calculations (see Fig.~\ref{fig:supFterms}(b)).
By following the idea of Vinet~\cite{vinet}, we consider the EOS at each temperature and fit the Helmholtz free energy-volume data $F(V)$ to various EOS models, including the Vinet and fourth-order Birch--Murnaghan model as introduced in the previous Sec.~\ref{subsec2:coldcurve}, as well as a numerical approach using cubic splines.
The B1-B2 transition pressures and volumes of the two phases upon transition can then be determined by the common tangents of $F(V)$ of the two phases (see Appendix~\ref{secapp:eosfit}).

\section{Results}\label{sec:result}
\subsection{Cold-curve equation of state}\label{subsec3:coldcurve}

The cold-curve EOS of B1 and B2 MgO based on static-lattice HF and AFQMC calculations are listed in Table~\ref{tab:afqmceos}.
The data for each phase at every volume is based on calculations using basis sets and simulation cells with finite sizes, which have then been extrapolated to the thermodynamic and CBS limits.
The results show that, for both B1 and B2 phases, the energy minimum locates at 17.0 to 18.7~\AA$^3$ when the calculation takes into account only exchange interactions of the electrons ($E_\text{HF}$), the correlation energy is about $-0.60$~Ha (1 Ha=27.211386 eV) at above 10.5~\AA$^3$ and decreases to $-0.63$~eV as the cell volume shrinks to $\sim$7~\AA$^3$, and the standard errors of the AFQMC data are small ($\sim$0.1 mHa).

\begin{table}[h]
\caption{\label{tab:afqmceos} Hartree--Fock ($E_\text{HF}$) and correlation ($E_\text{correlation}=E_\text{AFQMC}-E_\text{HF}$) energies of MgO in B1 and B2 phases at a series of volumes. The data are in the thermodynamic and CBS limits. $\sigma$ denotes the standard error of the AFQMC energy. Numbers are in units per Mg-O pair.}
    \centering
    \scriptsize
    \begin{ruledtabular}
    \begin{tabular}{rccc}
       $V$ (\AA$^3$) & $E_\text{HF}$ (Ha) & $E_\text{correlation}$  (Ha) & $\sigma$ (Ha) \\
         \hline
         B1 \\
7.2559 & -69.17120 & -0.62318 & 0.00014 \\
8.1877 & -69.37202 & -0.61553 & 0.00012 \\
9.1961 & -69.51994 & -0.60913 & 0.00012 \\
10.2840 & -69.62762 & -0.60393 & 0.00013 \\
11.4546 & -69.70457 & -0.59993 & 0.00013 \\
12.7107 & -69.75801 & -0.59671 & 0.00012 \\
14.0555 & -69.79346 & -0.59415 & 0.00027 \\
15.4919 & -69.81513 & -0.59388 & 0.00032 \\
17.0231 & -69.82624 & -0.59213 & 0.00015 \\
18.6519 & -69.82929 & -0.59054 & 0.00014 \\
20.3814 & -69.82618 & -0.59089 & 0.00019 \\
22.2147 & -69.81840 & -0.59072 & 0.00017 \\
         \hline  
         B2 \\
6.3976 & -68.99477 & -0.63091 & 0.00011 \\
7.2559 & -69.22190 & -0.62235 & 0.00011 \\
8.1877 & -69.39028 & -0.61602 & 0.00011 \\
9.1961 & -69.51385 & -0.61053 & 0.00012 \\
10.2840 & -69.60320 & -0.60677 & 0.00012 \\
11.4546 & -69.66637 & -0.60334 & 0.00015 \\
12.7107 & -69.70948 & -0.60184 & 0.00015 \\
14.0555 & -69.73719 & -0.59967 & 0.00016 \\
15.4919 & -69.75310 & -0.59856 & 0.00014 \\
17.0231 & -69.75998 & -0.59841 & 0.00015 \\
18.6519 & -69.75994 & -0.59909 & 0.00019 \\
20.3814 & -69.75465 & -0.60084 & 0.00030 \\
22.2147 & -69.74539 & -0.60147 & 0.00028 \\
    \end{tabular}
    \end{ruledtabular}
\end{table}

The energy-volume curves $E(V)$ are obtained
by fitting the AFQMC static lattice data to EOS models,
which gives rise to the equilibrium volume $V_0$ and bulk modulus $B_0$ of each phase.
The results are summarized in Table~\ref{tab:qmcvsdft} and compared 
to those from HF and DFT simulations in order to investigate the importance of the xc functionals.
We then calculated the B1-B2 transition pressure $P_\text{tr}$ and volumes of the two phases upon transition $V_\text{tr}$ from the common tangent of the $E(V)$ curves.
This is equivalent to another common approach for determining the transition pressure using the enthalpy-pressure relation (see Appendix~\ref{secapp:eosfit}).
Our results show that DFT predictions vary by up to $\sim$7\% in $V_0$, $\sim$15\% in $B_0$, $\sim$7\% in $P_\text{tr}$, and $\sim$10\% in volume change upon B1-B2 transition, due to usage of different xc functionals.

\begin{table*}[h]
\caption{\label{tab:qmcvsdft} Equilibrium volume ($V_0$), bulk modulus ($B_0$), and volumes of transition ($V_\text{tr}$) of MgO in B1 and B2 phases, and the transition pressure ($P_\text{tr}$), determined from the fitting of the $E(V)$ data from static lattice calculations using HF, DFT with different xc functionals, and AFQMC. Fittings are based on the fourth-order Birch--Murnaghan EOS model unless specified. Volumes are in units per Mg-O pair.
{\color{black}Also listed for comparison are results from the latest DMC calculations, which agree with our AFQMC predictions, and experimental values of $P_\text{tr}$, which are not precise enough to constrain theoretical predictions.}
}
    \centering
    \scriptsize
    \begin{ruledtabular}
    \begin{tabular}{lccccccc}
       & $V_0^\text{B1}$ (\AA$^3$) & $B_0^\text{B1}$ (GPa) &
       $V_0^\text{B2}$ (\AA$^3$) & $B_0^\text{B2}$ (GPa) & $V_\text{tr}^\text{B1}$ (\AA$^3$) & $V_\text{tr}^\text{B2}$ (\AA$^3$) & $P_\text{tr}$ (GPa)  \\
         \hline
         LDA & 18.054 & 172.1 & 17.676 & 158.8 & 8.849 & 8.445 & 531.7 \\
         PBE & 19.266 & 149.0 & 19.000 & 133.7 & 9.076 & 8.669 & 523.4 \\
         PBEsol & 18.737 & 157.4 & 18.366 & 145.1 & 9.013 & 8.609 & 517.5 \\
         SCAN & 18.469 & 166.2 & 19.720 & 75.3 & 8.914 & 8.483 & 546.5 \\
         SCAN~\footnotemark[1]\footnotetext{$E(V)$ fitted to Vinet EOS.} & 18.474 & 150.8 & 16.910 & 177.9 & 8.918 & 8.470 & 549.8 \\
         HSE06 & 18.564 & 166.7 & 18.130 & 153.9 & 8.983 & 8.568 & 530.6 \\
         HF~\footnotemark[2]\footnotetext{A different grid of (high-density only) data points is used for the B2 phase.} & 18.565 & 179.1 & 17.795 & 176.4 & 9.141 & 8.669 & 535.1 \\ 
         AFQMC~\footnotemark[2] & 18.407 & 175.7 & 17.940 & 154.6 & 9.201 & 8.739 & 499.2 \\
         {\color{black} DMC}~\footnotemark[3]\footnotetext{{\color{black}Diffusion Monte Carlo data (plus zero-point energy) fitted to a Vinet EOS by L. Shulenburger {\it et al.}~\cite{Root2015mgo}}} & {\color{black}18.788 $\pm$ 0.093} & {\color{black}153.8 $\pm$ 4.5} & -- & -- & -- & -- & {\color{black}493 $\pm$ 8 (503 $\pm$ 8}~\footnotemark[4]\footnotetext{{\color{black}Corrected to static lattice at 0 K.}})  \\ 
         {\color{black} Expt.}~\footnotemark[5]\footnotetext{{\color{black}Double-stage DAC measurements at room temperature by N. Dubrovinskaia {\it et al.}~\cite{Dubrovinskaia2019}}} & -- & -- & -- & -- & -- & -- & {\color{black}429--562 (439--572}~\footnotemark[6]\footnotetext{{\color{black}Corrected to static lattice at 0 K.}})  \\
         {\color{black}Expt.}~\footnotemark[7]\footnotetext{{\color{black}Laser-driven ramp compression at 2000--6000 K by F. Coppari {\it et al.}~\cite{Coppari2013}}} & -- & -- & -- & -- & -- & -- & {\color{black}410--600 }\\
    \end{tabular}
    \end{ruledtabular}
\end{table*}

\begin{figure*}[ht]
\centering
\includegraphics[width=0.9\linewidth]{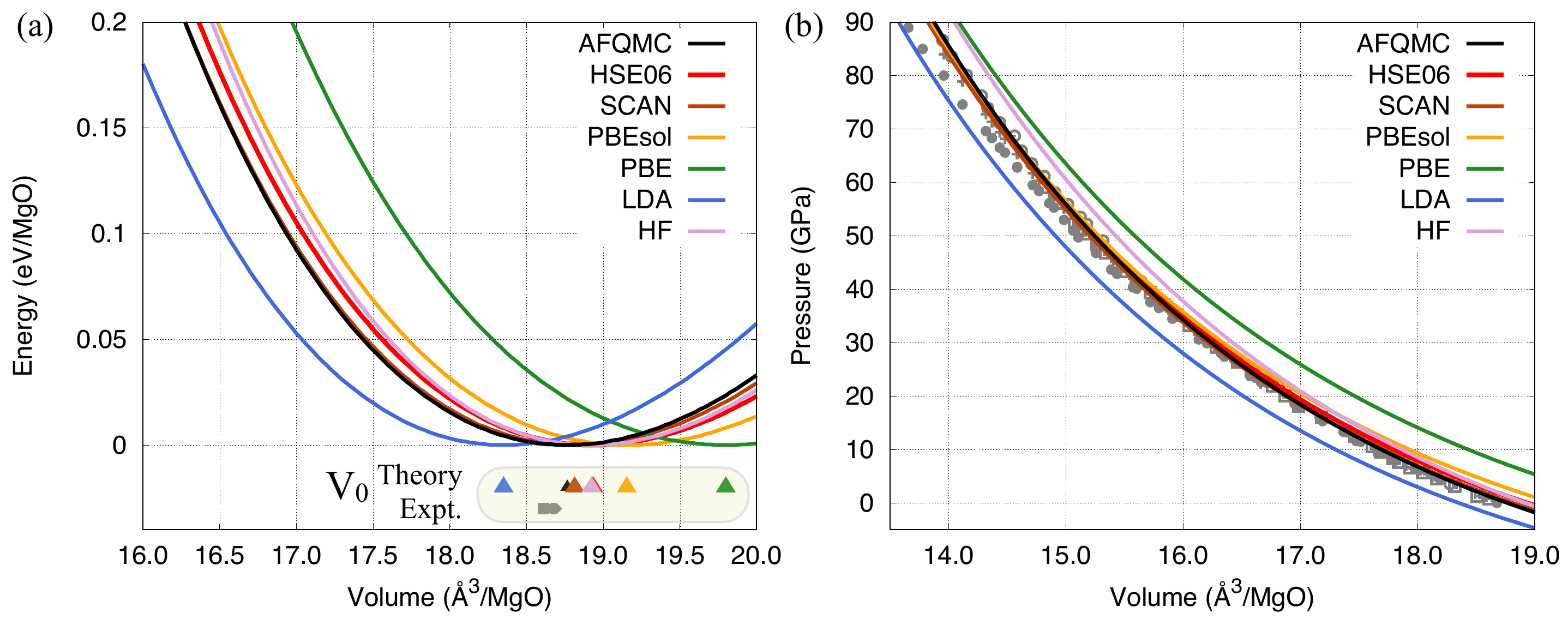}
\caption{AFQMC, various DFT, and HF predictions of (a) the energy-volume relation and equilibrium volume $V_0$ (colored triangles) and (b) compression curve of B1 MgO at 300 K, benchmarked against experimental values (gray symbols) from Refs.~\onlinecite{Fei1999,Speziale2001,Jacobsen2008}. In (a), all energy curves are plotted relative to their respective minimum values.}
\label{fig:benchmarkB1}
\end{figure*}

To directly compare theoretical EOS with DAC experiments, corrections to the static-lattice results are needed to account for the differences due to lattice vibration and thermal contributions.
We have added ion thermal contributions to the cold curve EOS via lattice vibration calculations under QHA,
which is generally considered a good approximation for MgO under room temperature.
The cold-curve EOS is re-evaluated by fitting the corrected 300-K data for each phase to the EOS models.
The equilibrium volume and the pressure-volume results from theoretical calculations (AFQMC, DFT, and HF) are shown in Fig.~\ref{fig:benchmarkB1} and compared to experimental results.
It shows remarkable agreement between AFQMC and experimental results for both the equilibrium volume and compression curve.
In contrast, the HF and DFT results scatter around the experimental values and vary significantly.
Explicitly, DFT results exhibit a strong dependence on the choice of the xc functional,
with HSE06, SCAN, and PBEsol performing better than PBE and LDA when compared with the experimental data.

\begin{figure*}[ht]
\centering
\includegraphics[width=0.9\linewidth]{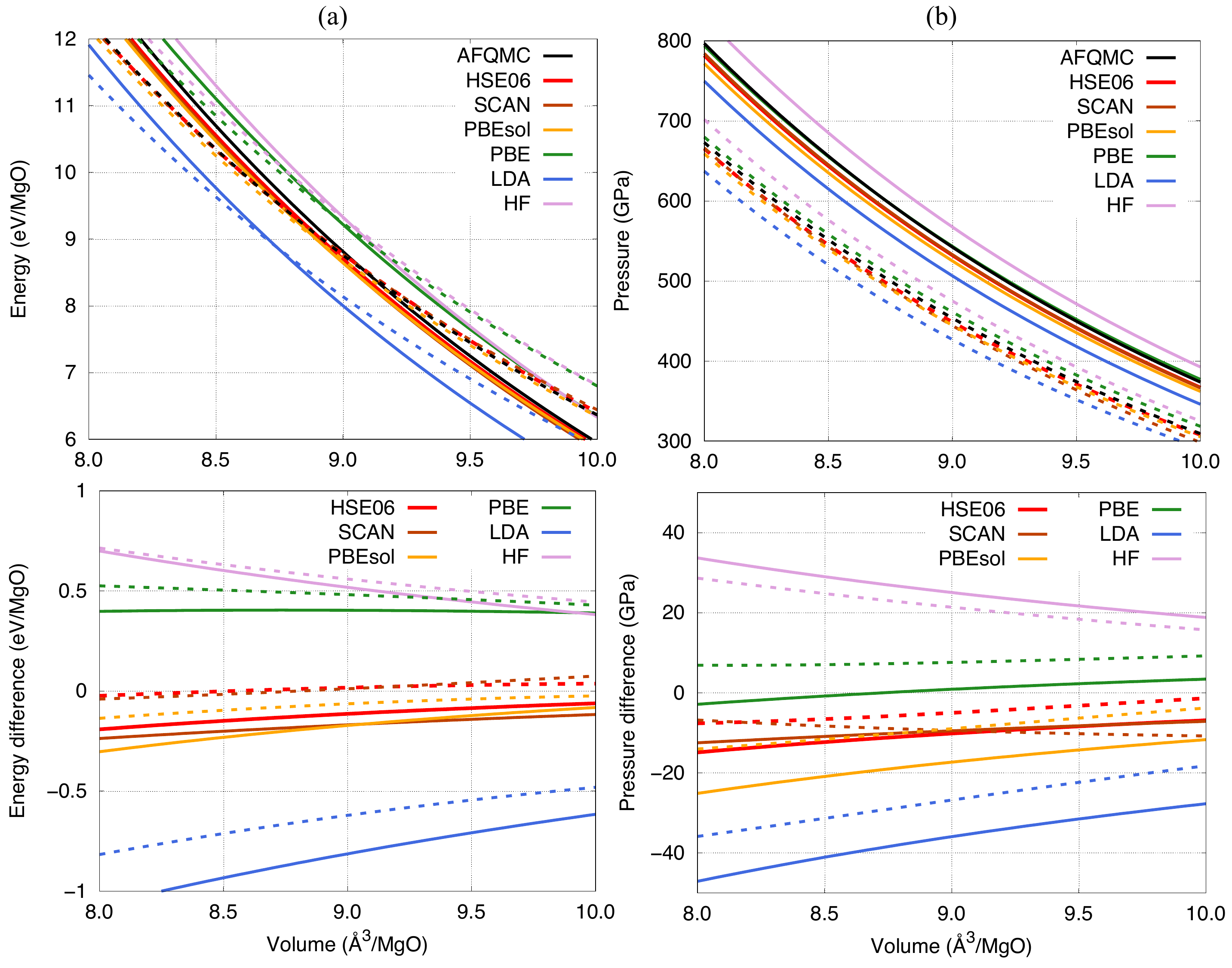}
\caption{Comparison between AFQMC and various DFT or HF predictions of (a) internal energies and (b) pressures from static-lattice calculations of MgO around the B1-B2 transition. Top: direct comparisons; bottom: differences relative to AFQMC. Solid and dashed lines denote results for the B1 and B2 structures, respectively.}
\label{fig:benchmarkXC}
\end{figure*}

Figure~\ref{fig:benchmarkXC}
compares the AFQMC cold curve of MgO at higher densities (near the B1-B2 transition) with those calculated using HF and DFT with various xc functionals.
It shows HF, LDA and PBE demonstrate large deviations (approximately $\pm$0.5 to 1 eV/MgO in energy and 0 to 40 GPa in pressure, depending on the phase and the density) for the cold curves, while PBEsol, SCAN, and HSE06 show significantly improved agreements, in comparison to the AFQMC results.
These findings are overall consistent with normal expectations based on Jacob's ladder (precision relation: hybrid$>$meta-GGA$>$GGA/LDA$>$HF).

\begin{figure}[ht]
\centering
\includegraphics[width=0.7\linewidth]{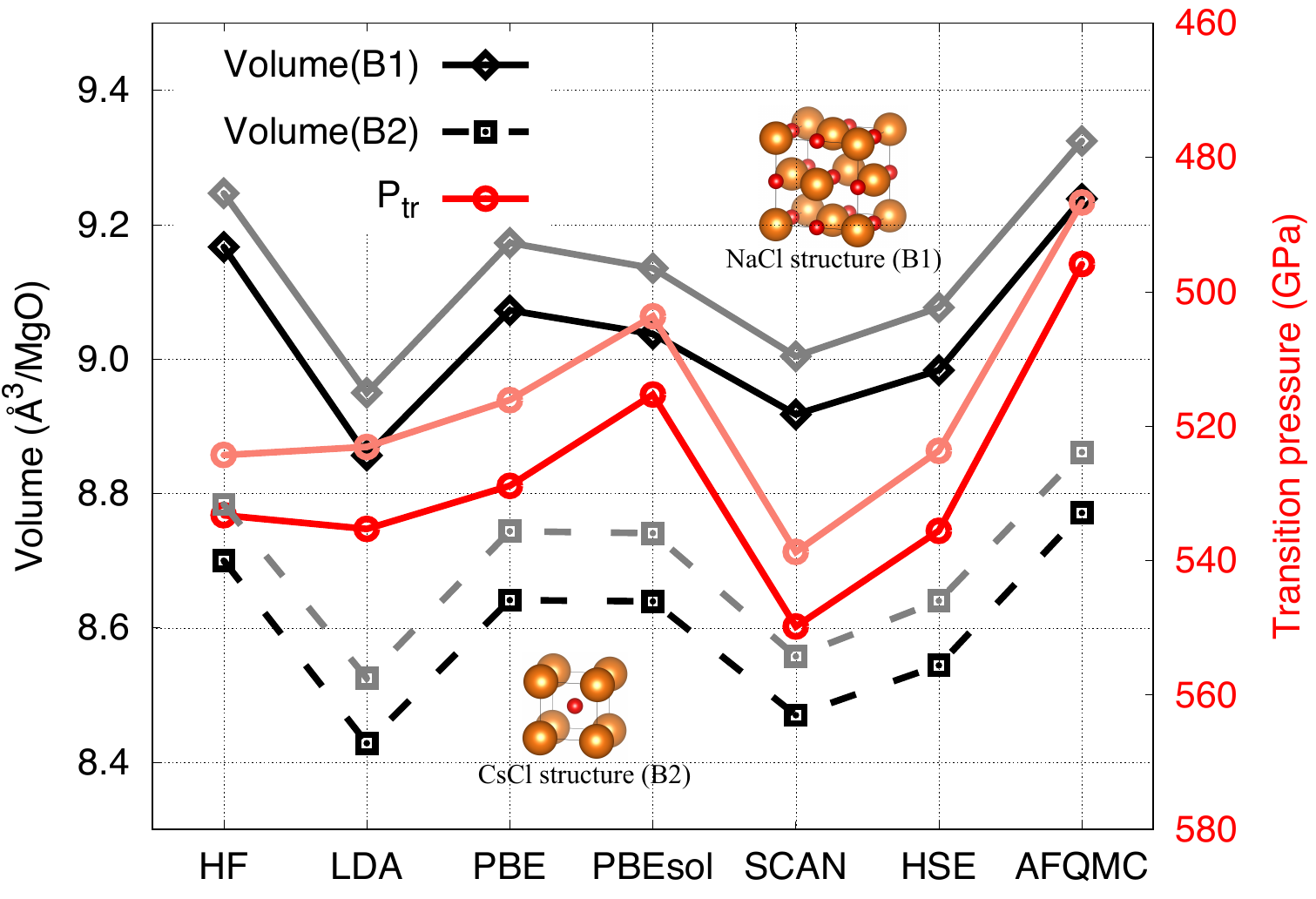}
\caption{Comparison between AFQMC and various DFT or HF predictions of the volumes of the B1 and B2 phases upon transition and the transition pressure. Lighter colors denote 300-K results based on QHA; darker colors denote static-lattice results.
The transition occurs at 429 to 562~GPa according to room-temperature experiments~\cite{Dubrovinskaia2009}.}
\label{fig:trans0K}
\end{figure}

Figure~\ref{fig:trans0K} summarizes the B1-B2 phase transition pressures (red) and volume changes upon the phase transition (black) of MgO calculated using HF and DFT with various XC functionals in comparison to AFQMC. 
Due to the reconciliation of the EOS errors for the B1 and B2 phases in the calculation of the B1-B2 transition,
the proximity of HF and DFT to AFQMC results no longer follows expectations of Jacob's ladder.
The AFQMC predicted transition pressure is lower and volumes upon transition are larger than all other methods.
PBEsol prediction of the transition pressure is closer to AFQMC than HF or other DFT xc functionals, with a difference of 20 GPa.

\subsection{High-temperature EOS}\label{subsec3:highT}

High-temperature EOS of solid-state MgO is obtained from QMD and QHA calculations.
The QHA results are based on a combination of phonon and cold-curve EOS, where the cold curves are obtained by static DFT calculations using four different xc functionals (LDA, PBE, PBEsol, and SCAN), while phonon calculations are performed by using the DFPT approach, PBEsol xc functional, Mg\_sv\_GW and O\_h pseudopotentials, and including the Born effective charge. Tests show negligible differences in vibrational energies if the phonon calculations are done by using other xc functionals or ignoring the splitting between longitudinal and transverse optical modes (see Appendices~\ref{secapp:loto} and~\ref{secapp:xc}).




\begin{figure*}[ht]
\centering
\includegraphics[width=1.0\linewidth]{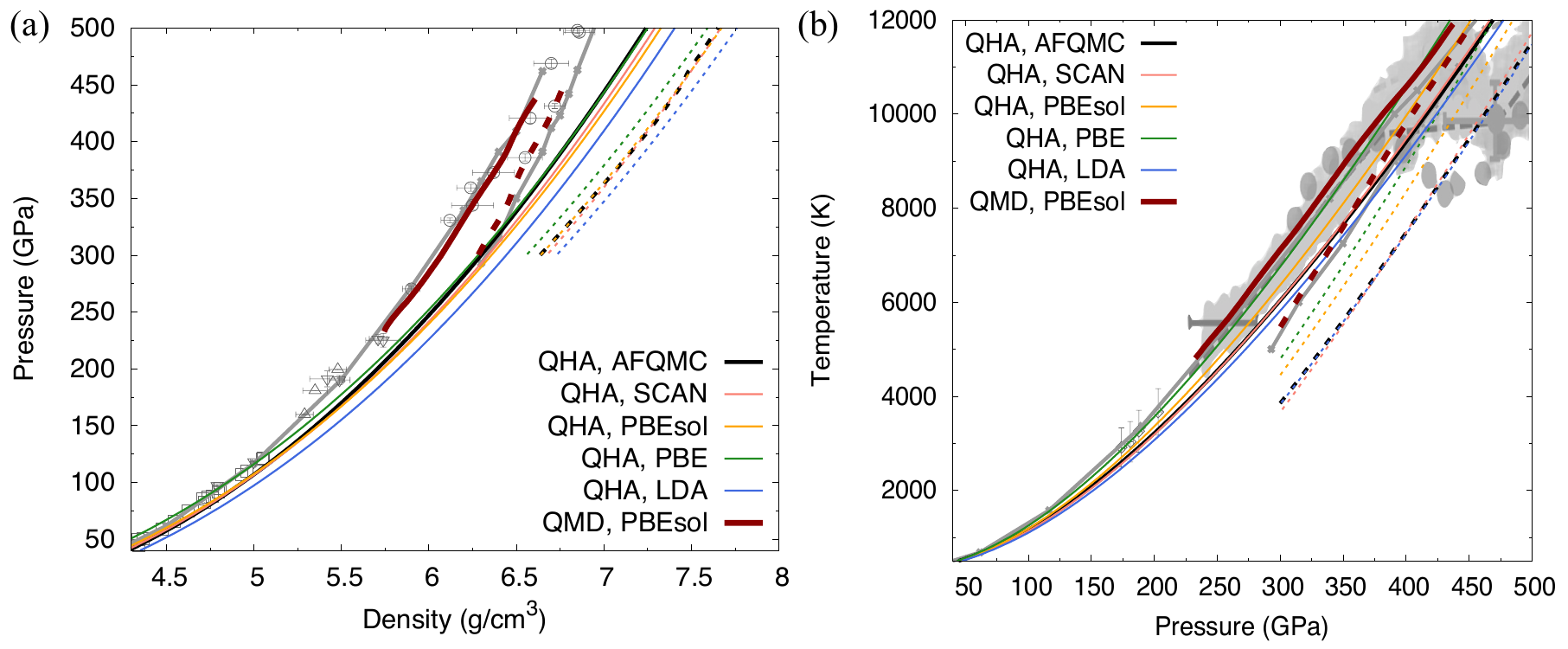}
\caption{Comparison between QMD and QHA Hugoniots of MgO in (a) pressure-density and (b) temperature-pressure spaces. 
Different colors denote the different calculation methods.
Solid and dashed curves represent results for B1 and B2 phases, respectively. 
Also included (in gray) are the experimental Hugoniot data from Marsh {\it et al.} (squares)~\cite{MarshLASL1980}, Vassiliou {\it et al.} (up triangles)~\cite{Vassiliou1981}, Fratanduono {\it et al.} (down triangles)~\cite{Fratanduono2013mgo}, Root {\it et al.} (circles)~\cite{Root2015mgo}, Svendsen and Ahrens {\it et al.} (diamonds)~\cite{Svendsen1987}, McWilliams {\it et al.} (ovals)~\cite{McWilliamsScience2012}, and Bolis {\it et al.} (shaded areas, with a horizontal bar denoting the error in pressure and a cross denoting the condition interpreted as the melting point)~\cite{Bolis2016mgo}, previous DFT-MD predictions based on the Armiento--Mattsson xc functional (line-crosses)~\cite{Root2015mgo}, and a thermodynamic EOS model by de Koker and Stixrude (thick dashed lines)~\cite{dKS2009}. 
}
\label{fig:hugQMDvsQHAexpt}
\end{figure*}

We first used the EOS to calculate the principal Hugoniot and compared them with experiments. 
Figure~\ref{fig:hugQMDvsQHAexpt} shows comparisons of the Hugoniots in pressure--density and temperature--pressure spaces.
Similar to previous QMD calculations that used the Armiento--Mattsson (AM05) xc functional~\cite{Root2015mgo}, our present QMD results based on the PBEsol functional show excellent agreement with experimental Hugoniots in stability regimes of both B1 and B2 in the pressure-density relation, as well as for B1 in the temperature profile.
In comparison, QHA results show consistency with experiments at low pressures but give increasingly higher density at high pressures along the Hugoniot, more so for the B2 than the B1 phase.
The breakdown of QHA as shown in the pressure-density results can be attributed to the anharmonic vibration effect that is naturally included in QMD but missing in QHA and becomes more significant at higher temperatures.
By comparing the thermal EOS along an isotherm, we found similar energies but higher pressures given by QMD than by QHA;
according to the Rankine--Hugoniot equation, this must be reconciled by less shrinking in volume,
which explains the Hugoniot density relations between QMD and QHA as shown in Fig.~\ref{fig:hugQMDvsQHAexpt}(a).

In the temperature-pressure space, QMD and QHA results of the Hugoniot are less distinct from each other than in the pressure-density space. 
QHA results based on LDA xc functional clearly lie below the range of the experimental data for the B1 phase, PBE significantly improves the agreement with experiments, while PBEsol and SCAN functionals and AFQMC data fall between LDA and PBE and near the lower bound of the experimental data.
QMD predictions of the temperature are higher and improve over that by QHA using PBEsol.
In addition, QMD predicts smaller differences between the Hugoniot of the B1 and B2 phases than QHA;
AFQMC predictions of the B2 Hugoniot show good agreement with SCAN and LDA under QHA, following the trend of experimental Hugoniot after the turnover, while the QHA-PBEsol predictions are slightly higher.
Our QMD results of the Hugoniot are overall consistent with previous calculations and {\color{black}align with experiments.
More discussions will be given in the following and in Appendix~\ref{secapp:PtrvsExpt} regarding the B1--B2 phase boundary and comparison between our prediction and the experiments.}

The agreement with experiments in both the 
thermal (along the Hugoniot) and the cold-curve EOS (as shown in the previous Sec.~\ref{subsec3:coldcurve}) validates PBEsol as an optimal choice for the xc functional for calculations of MgO at both the ground state and finite temperatures. 
In the following, we have added the QHA and QMD-derived (using the TDI approach) thermal free energies based on DFT-PBEsol calculations to various cold curves (by AFQMC and DFT-PBEsol/SCAN) to estimate the total free energies of MgO in both B1 and B2 phases~\footnote{We encounter imaginary-mode problems when using SCAN for phonon calculations at some of the densities.
Therefore, we only use PBEsol for the QHA and finite-temperature QMD calculations.}.
Based on these results, we charted the B1-B2 transition and calculate the volumes of the two phases upon transition.
The results provide a preliminary reference for the B1-B2 phase boundary and its uncertainty based on state-of-art theoretical computations.

\begin{figure*}[ht]
\centering
\includegraphics[width=0.85\linewidth]{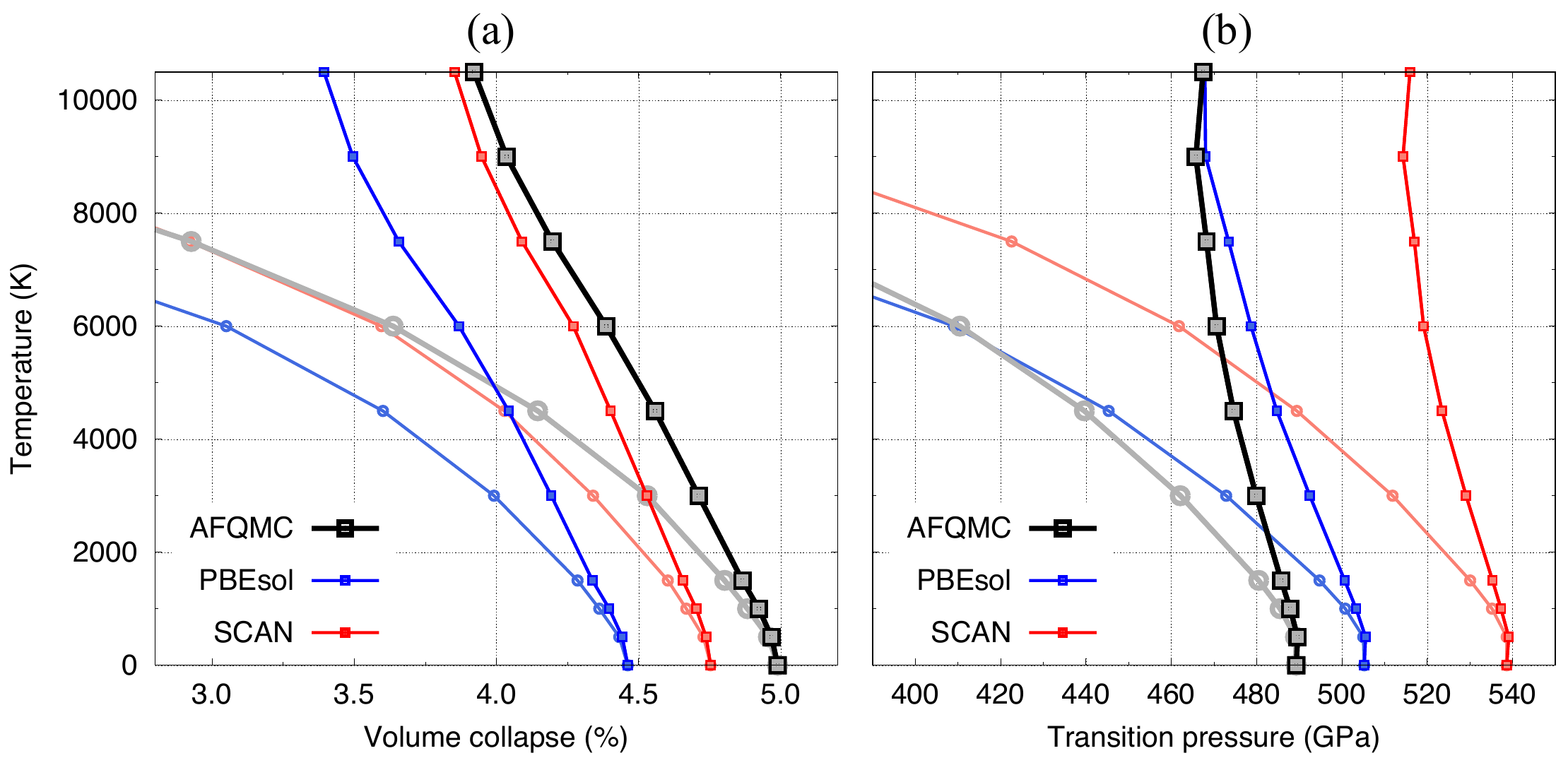}
\caption{ (a) Volume collapse and (b) pressures of the B1$\rightarrow$B2 transition at finite temperatures with the cold curve calculated using AFQMC and DFT with two optimal xc functionals (PBEsol and SCAN) and thermal effect (based on QMD and TDI) calculated within DFT using the PBEsol functional.
Results based on QHA (including electron thermal but excluding anharmonic vibrational effects) are shown in lighter-colored curves (with circle symbols) for comparison.
}
\label{fig:B1B2trans}
\end{figure*}

Figure~\ref{fig:B1B2trans} shows the volume of MgO
collapses by $\sim4.75(\pm0.25)\%$ at 0 K [from $\sim$9.2 \AA$^3$/MgO for B1 to $\sim$8.7 \AA$^3$/MgO for B2 (error associated with {\color{black}using} different methods {\color{black} AFQMC, PBEsol, and SCAN}: $\pm$0.1 \AA$^3$/MgO)] and $3.7(\pm0.2)\%$ at 10,500 K [from $\sim$9.8 \AA$^3$/MgO for B1 to $\sim$9.4 \AA$^3$/MgO for B2 (error: $\pm$0.2 \AA$^3$/MgO)] for the B1$\rightarrow$B2 transition, and
the transition pressure decreases from $\sim515(\pm25)$ GPa to $\sim490(\pm25)$ GPa as temperature increases from 0 to 10,500 K.
We found the $V_\text{tr}-T$ curves are similar between the three sets of predictions based on AFQMC and DFT-PBEsol/SCAN cold curves,
with the AFQMC predicted volumes and volume collapses larger (and transition pressures lower) than the DFT predictions.
The $\mathrm{d}T/\mathrm{d}P$ Clapeyron slope of the B1-B2 phase boundary predicted by the AFQMC data set is similar to DFT-SCAN, both being steeper than that by DFT-PBEsol {\color{black}(see Table~\ref{tab:dPdTtr} that summarizes values of the Clapeyron slope)}.

Figure~\ref{fig:B1B2trans} also shows QHA predicts a much less steep boundary for the B1-B2 transition than QMD, reflecting the importance of anharmonic vibrational effects, similar to the report by previous studies~\cite{Bouchet2019,Soubiran2020}.
Our results clearly show the amounts of changes in volume and volume difference between the two phases of MgO upon transition, as well as the important role of electronic interactions (many-body in nature {\color{black}versus single-particle approximation} under different xc functionals) in affecting the results. 
{\color{black}The much less negative value in Clapeyron slope ($dP_\mathrm{tr}/dT$) and slightly larger value in volume collapse of the B1--B2 transition predicted by QMD may 
cause less significant topography of discontinuity and lateral variations in deep-mantle mineralogy of super-Earths than previously expected based on the QHA results, changing expectations on the style of convection in these planets (see discussions in, e.g., ~\cite{KaratoBook2003}).

Moreover, by predicting a steeper B1--B2 boundary than latest theoretical studies~\cite{Bouchet2019,Soubiran2020}, our AFQMC (and PBEsol) results show excellent consistency with both experiments by McWilliams {\it et al}~\cite{McWilliamsScience2012} and Bolis {\it et al}~\cite{Bolis2016mgo} (see Appendix~\ref{secapp:PtrvsExpt}).
We note that Bolis {\it et al}~\cite{Bolis2016mgo} interpreted the turnover in their experiments as the melting start of shocked MgO, largely based on comparisons with theoretical studies by then that underestimated $P_{\mathrm{tr}}$ along the Hugoniot. Our new results suggest that the turnovers in the experiments are associated with the B1–B2 transition.
It is beyond the scope of this study, however, to decipher the nature of the subtle differences between experiments by Bolis {\it et al}.~\cite{Bolis2016mgo} and McWilliams {\it et al}.~\cite{McWilliamsScience2012},
as it requires accurate knowledges of the triple point, thermodynamic free energies of the liquid phase, as well as considerations of the kinetics of the transition to fully understand the observations.
}

We have performed additional tests and found the error in transition pressure (associated with the choices of different $T_\text{ref}$ and fitting methods in TDI) increases to $\sim$50~GPa at $T\approx10^4$~K (see Appendix~\ref{secapp:Fah}, {\color{black}corresponding changes in the Clapeyron slope are tabulated in Table~\ref{tab:dPdTtr}}),
while the errors due to other sources (EOS models, data error bars, and the data grid) are relatively small (e.g.,
the statistical error of the AFQMC and QMD energies only leads to a difference in $P_\text{tr}$ of 1.5 GPa at 6000 K).


\section{Conclusions}\label{sec:conclusions}
This work exemplifies the first application of the AFQMC approach to benchmark the cold curve and phase transition in solid-state materials to very high pressures.
Our AFQMC results reproduce the experimental cold curve (equilibrium volume and compressibility at room temperature) and
provide a preliminary reference for the equations of state of MgO at up to 1 TPa.
In comparison, DFT predictions vary by up to 7\% to 15\% for the equilibrium properties ($V_0$ and $B_0$) and B1-B2 transition ($P_\text{tr}$ and volume collapse upon the transition), depending on the xc functionals, and the largest differences are observed between the cold curves by PBE and LDA.
The HSE06, SCAN, PBEsol functionals perform better than PBE, LDA, and HF in reproducing the $E(V)$ cold curves by AFQMC.
The cold-curve differences for B1 offset those for B2, leading to the sensitivity of the predicted transition pressure and volume change to the choice of the xc functional.

Our Hugoniot results based on QMD calculations of the thermal EOS using PBEsol show excellent agreement with experiments
for B1 and B2 in the pressure-density relation, as well as for B1 in the temperature-pressure profile.
In comparison, QHA results of the pressure-density Hugoniot show consistency with experiments at low pressures but increasing discrepancy at high pressures, because larger anharmonic effects are expected at higher temperatures.
The good performance of PBEsol in reproducing both the thermal (along the Hugoniot) and the cold-curve EOS of MgO has motivated us to further calculate the anharmonic free energies and add them to the cold curves by AFQMC and DFT-PBEsol or SCAN to calculate the total free energies and evaluate the B1-B2 transition at various temperatures.
Our results show temperature lowers the transition pressure and expands the volumes upon the B1-B2 transition.
Anharmonic vibration increases the transition pressure $P_\text{tr}$ and hinders the transition volumes $V_\text{tr}$ from expansion, relative to QHA.
AFQMC predicts a steeper $\mathrm{d}T/\mathrm{d}P$ phase boundary and a larger volume collapse upon the B1$\rightarrow$B2 transition than DFT-PBEsol, similar to the effect of anharmonicity with respect to QHA.

In addition to providing a preliminary reference
for the B1-B2 phase boundary and its uncertainty
based on state-of-art theoretical computations,
our results will be useful for 
building an accurate multiphase EOS table for MgO for planetary sciences and high energy density sciences applications, as well as for elucidating the mechanism of phase transformations (e.g., kinetics effects) in different experimental settings (e.g., compression rates).
More work is desired to clarify the triple point and the melting curve at high temperatures and pressures to multi-TPa pressures. 
Looking ahead, finite-temperature AFQMC~\cite{Lee2021jcp,Shen2020jcp}, by better accounting of the electron thermal effects, {\color{black}and back-propagation for force and stress estimators in AFQMC~\cite{chen2023computation} can offer additional yet more-accurate options to benchmark the EOS and phase transitions of solid-state materials} at high temperatures and pressures.

\section*{Acknowledgements}
This material is based upon work supported by the Department of Energy National Nuclear Security Administration under Award Number DE-NA0003856, the University of Rochester, and the New York State Energy Research and Development Authority.
The Flatiron Institute is a division of the Simons Foundation.
Part of this work was performed under the auspices of the U.S. Department of Energy by Lawrence Livermore National Laboratory under contract number DE-AC52-07NA27344. 
Part of the funding support was from the U.S. DOE, Office of Science, Basic Energy Sciences, Materials Sciences and Engineering Division, as part of the Computational Materials Sciences Program and Center for Predictive Simulation of Functional Materials (CPSFM). 
Computer time was provided by the Oak Ridge Leadership Computing Facility, Livermore Computing Facilities, and UR/LLE HPC.
{\color{black}S.Z. thanks R. S. McWilliams for sharing experimental data and J. Wu, R. Jeanloz, F. Soubiran, B. Militzer, T. Duffy, and K. Driver for discussions.}

This report was prepared as an account of work sponsored by an agency of the U.S. Government. Neither the U.S. Government nor any agency thereof, nor any of their employees, makes any warranty, express or implied, or assumes any legal liability or responsibility for the accuracy, completeness, or usefulness of any information, apparatus, product, or process disclosed, or represents that its use would not infringe privately owned rights. Reference herein to any specific commercial product, process, or service by trade name, trademark, manufacturer, or otherwise does not necessarily constitute or imply its endorsement, recommendation, or favoring by the U.S. Government or any agency thereof. The views and opinions of authors expressed herein do not necessarily state or reflect those of the U.S. Government or any agency thereof. 


\appendix

{\color{black}
\section{Finite-size and basis sets corrections to AFQMC energies}\label{secapp:afqmcconverge}
}

\begin{figure*}[ht]
\centering
\includegraphics[width=0.65\linewidth]{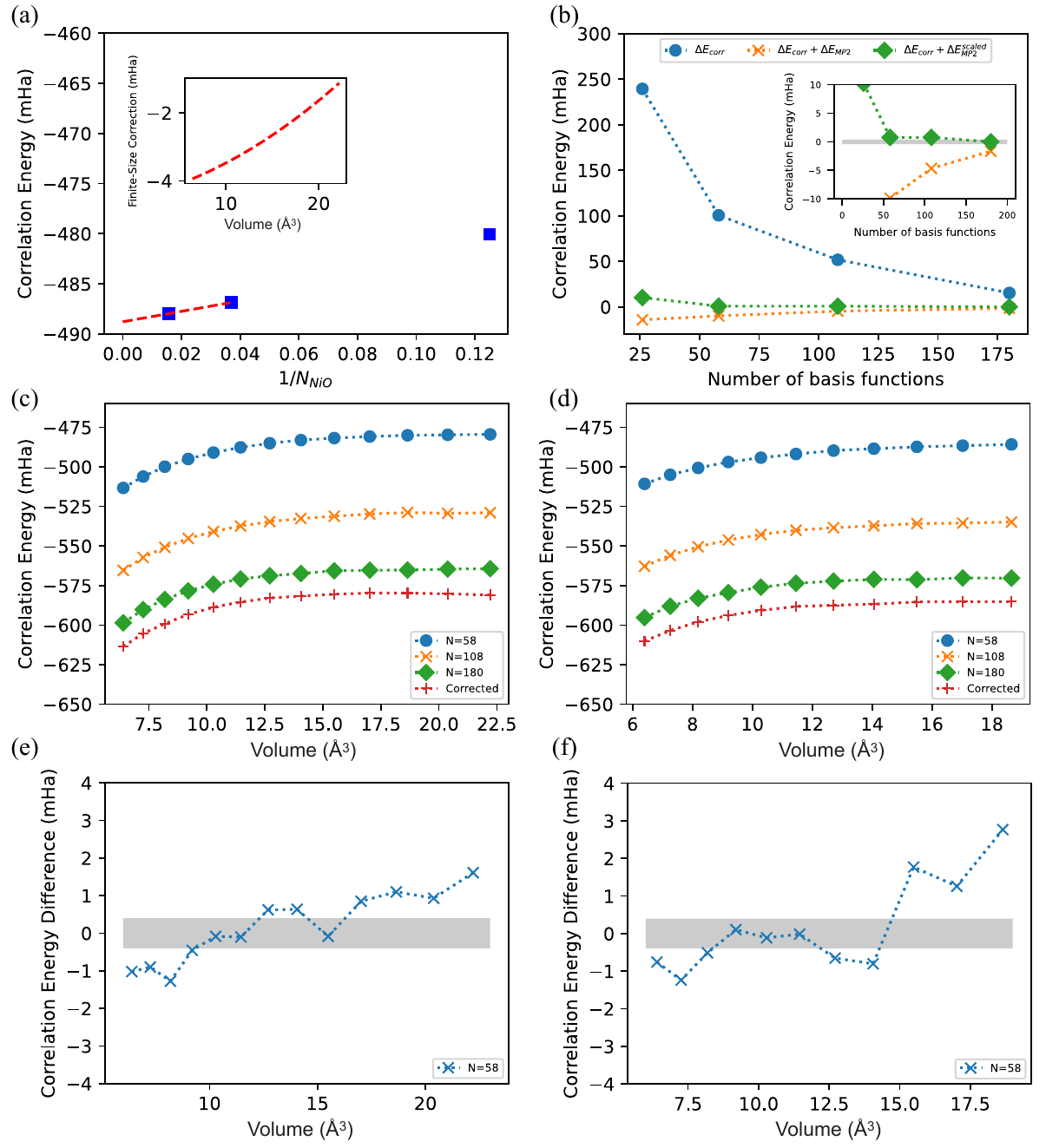}
{\color{black}
\caption{Examples of our AFQMC energies extrapolation: (a) finite-size correction [pVTZ ($N$=58)]; (b) AFQMC correlation energy measured with respect to the CBS limit;
(c-d) uncorrected AFQMC correlation energies for various basis sets and, as a comparison, basis set corrected AFQMC energies with the pVTZ basis set; (e-f) the same red curves in panels (c-d) but now measured with respect to corrected pV5Z basis set results, showing the excellent agreement and efficiency of the basis set correction. Panels (a-b): 18.65~\AA$^3$/MgO; (b-f): $N_\mathrm{NiO}=8$; (a-c) and (e): B1; (d) and (f): B2. The inset of (a) shows the values of the finite-size corrections at different volumes. In (b), {\color{black} $\Delta E_\mathrm{corr}$ denotes uncorrected AFQMC energies,  $\Delta E_\mathrm{MP2}$ represents the MP2 correction, and $\Delta E_\mathrm{MP2}^\mathrm{scaled}$ denotes the scaled MP2 correction}. $N$ denotes the number of basis functions.}
\label{fig:AFQMCconverge}
}
\end{figure*}

{\color{black}
Our AFQMC calculations were performed for both phases of MgO at all volumes with various cell sizes and optimized basis sets. These include: 
(i) 2$\times$2$\times$2 $k$ points (8 MgO units) with pVDZ, pVTZ, pVQZ, and pV5Z; 
(ii) 3$\times$3$\times$3 $k$ points (27 MgO units) with pVTZ and pVQZ; and
(iii) 4$\times$4$\times$4 $k$ points (64 MgO units) with pVTZ.

We have then followed three steps to extrapolate the AFQMC results to the thermodynamic and the complete basis set (CBS) limits:\\
\indent 1. Use all the pVTZ results to calculate finite size corrections for the 3$\times$3$\times$3 calculations;\\
\indent 2. Use all the 3$\times$3$\times$3 calculations to calculate the basis set corrections, combining AFQMC calculations with MP2 calculations and the ``scaled'' correction described in Ref.~\onlinecite{Morales2020};\\
\indent 3. Use the 2$\times$2$\times$2 calculations to check reliability of the basis set corrections in step 2 and to ensure the basis set corrections were robust.

Our extrapolation procedure is demonstrated in Figure~\ref{fig:AFQMCconverge}. The remarkable consistency between the pVTZ and pV5Z corrected values (to approximately 1--2 mHa/MgO from calculations with only 2$\times$2$\times$2 $k$ points) suggests our corrections are reliable and robust.


}

\section{EOS fit and transition pressure determination}\label{secapp:eosfit}

\begin{figure*}[ht]
\centering
\includegraphics[width=0.9\linewidth]{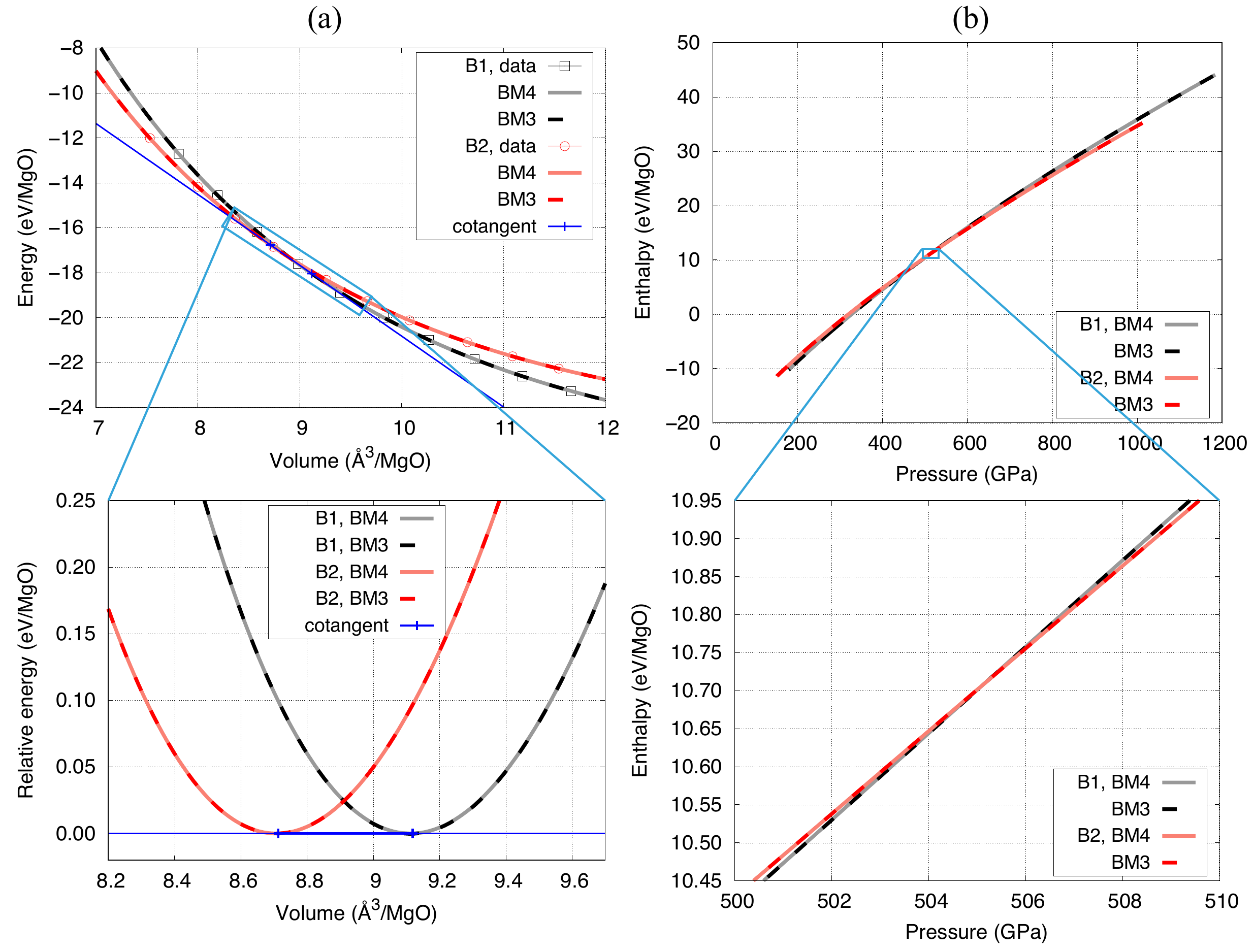}
\caption{Determination of the transition pressure ($P_\text{tr}$) by using two different approaches: (a) common-tangent of the internal energy $E(V)$ curves and (b) intersect of the enthalpy $H(P)$ curves. In this example, the data are from 0-K DFT-PBEsol+QHA calculations; $P_\text{tr}$ determined using the two approaches are 504 and 505 GPa, respectively.}
\label{fig:Ptr}
\end{figure*}

Figure~\ref{fig:Ptr} compares the two different ways of calculating the transition pressure: using internal energies $E(V)$ and their common tangent (left) or using enthalpies $H(P)$ and their crossover point~\footnote{This example is for $T$=0 K. At finite temperatures, we consider isotherms and use Helmholtz free energies $F(V)$ for the common-tangent approach or Gibbs free energies $G(P)$ for the crossover-point approach.}.
The two approaches are thermodynamically equivalent, as shown by the same transition pressure that has been determined (the 1-GPa difference is due to the numerical fitting of the data).
The common-tangent approach is our option in this study because the internal energy (or Helmholtz free energy for $T\neq$ 0 K) is readily calculated while pressure is not except for the 0-K DFT cases.

\begin{figure}[ht]
\centering
\includegraphics[width=0.7\linewidth]{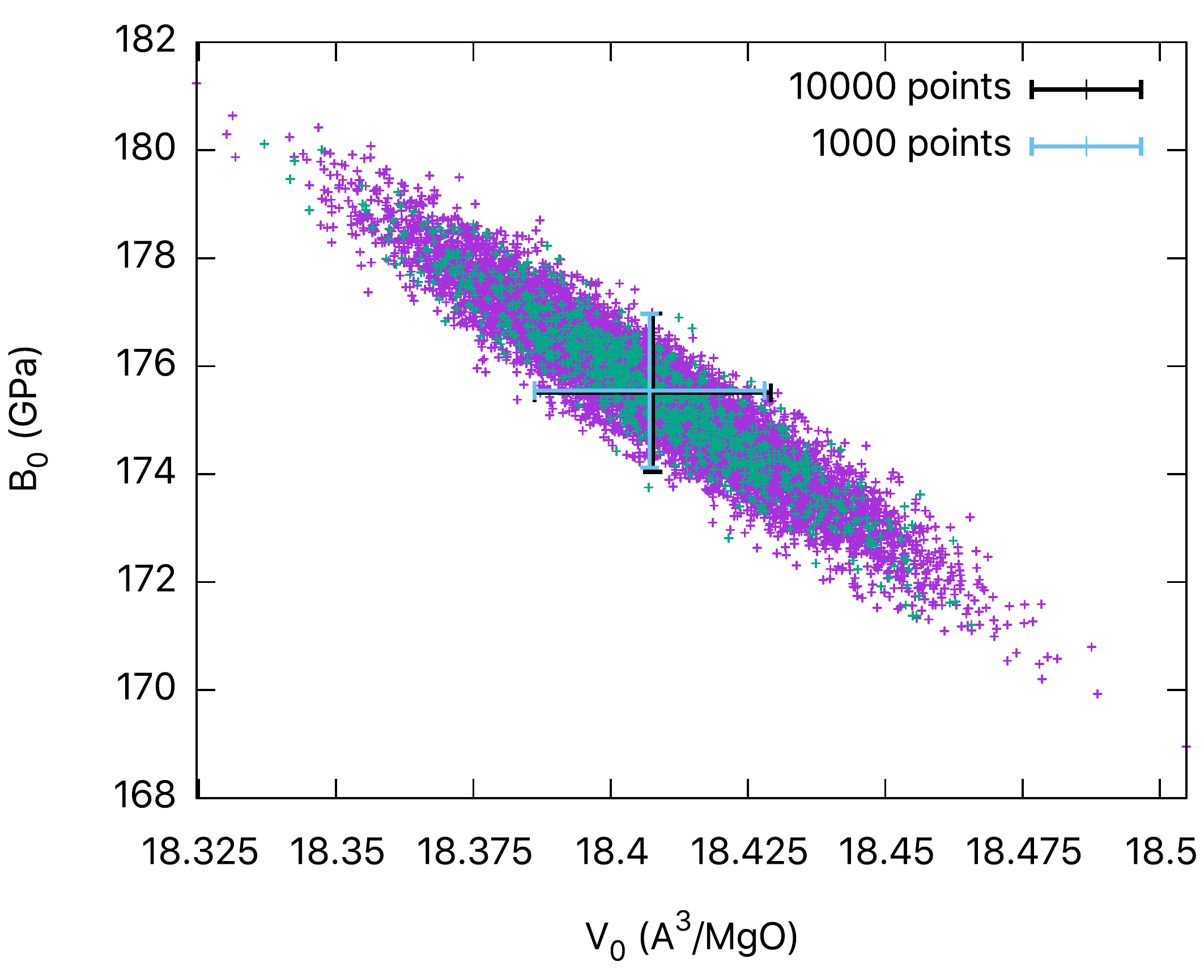}
\caption{A Monte Carlo approach is used to determine the 1-$\sigma$ errors of the EOS fitting parameters. In this example,
the AFQMC data and their standard errors (as shown in Table~\ref{tab:afqmceos}) are used and the third-order Birch--Murnaghan EOS model is considered. Two Monte Carlo runs with 1000 or 10000 randomly generates datasets give the same results for $V_0$, $B_0$, and their error bars.}
\label{fig:fiterror}
\end{figure}

The $E(V)$ data are fitted to EOS models to determine the equilibrium volume $V_0$ and bulk modulus $B_0$. 
Typical errors of these parameters can be calculated using the Monte Carlo approach and are shown in Fig.~\ref{fig:fiterror}.

Table~\ref{tab:appen2} summaries the equilibrium volume $V_0$ and bulk modulus $B_0$ 
by using different EOS models for the PBEsol data.
We found the third-order Eulerian EOS (Birch--Murnaghan) works surprisingly well for MgO up to TPa pressures as long as data at high-enough pressures are included.

\begin{table*}[h]
\caption{\label{tab:appen2} Equilibrium volume ($V_0$), bulk modulus ($B_0$) and its pressure derivative ($B_0'$), and volumes of transition ($V_\text{tr}$) of MgO in B1 and B2 phases, and the transition pressure ($P_\text{tr}$), determined in different fitting approaches for the $E(V)$ data from static DFT calculations using the PBEsol xc functional. Volumes are in units per Mg-O pair.}
    \centering
    \scriptsize
    \begin{ruledtabular}
    \begin{tabular}{lccccccccc}
       & $V_0^\text{B1}$ (\AA$^3$) & $B_0^\text{B1}$ (GPa) & $B_0^{\prime\text{B1}}$ &
       $V_0^\text{B2}$ (\AA$^3$) & $B_0^\text{B2}$ (GPa) & $B_0^{\prime\text{B2}}$ & 
       $V_\text{tr}^\text{B1}$ (\AA$^3$) & $V_\text{tr}^\text{B2}$ (\AA$^3$) & $P_\text{tr}$ (GPa)  \\
         \hline
         BM3 & 18.720 & 161.4 & 4.01 & 17.797 & 170.2 & 3.94 & -- & -- & -- \\
         BM4 & 18.737 & 157.4 & 4.11 & 18.366 & 145.1 & 4.09 & 9.013 & 8.609 & 517.5 \\
         Vinet & 18.786 & 137.7 & 4.91 & 20.552 & 68.5 & 5.56 & 9.012 & 8.583 & 522.7 \\
         BM3~\footnotemark[1] & 18.721 & 160.4 & 4.02 & 18.266 & 151.3 & 3.99 & -- & -- & -- \\
         BM4~\footnotemark[1] & 18.737 & 157.3 & 4.11 & 18.289 & 147.6 & 4.09 & 9.014 & 8.610 & 517.4 \\
         Vinet~\footnotemark[1] & 18.791 & 141.1 & 4.81 & 18.364 & 129.2 & 4.86 & 9.037 & 8.639 & 515.3 \\
         spline~\footnotemark[1] & -- & -- & -- & -- & -- & -- & 9.001 & 8.605 & 518.5 \\
    \end{tabular}
    \end{ruledtabular}
    \footnotetext{Different grids of data points (slightly denser for B1 and high-density only for B2) are used.}
\end{table*}

\section{Effect of LO-TO splitting}\label{secapp:loto}
\begin{figure*}[ht]
\centering
\includegraphics[width=0.8\linewidth]{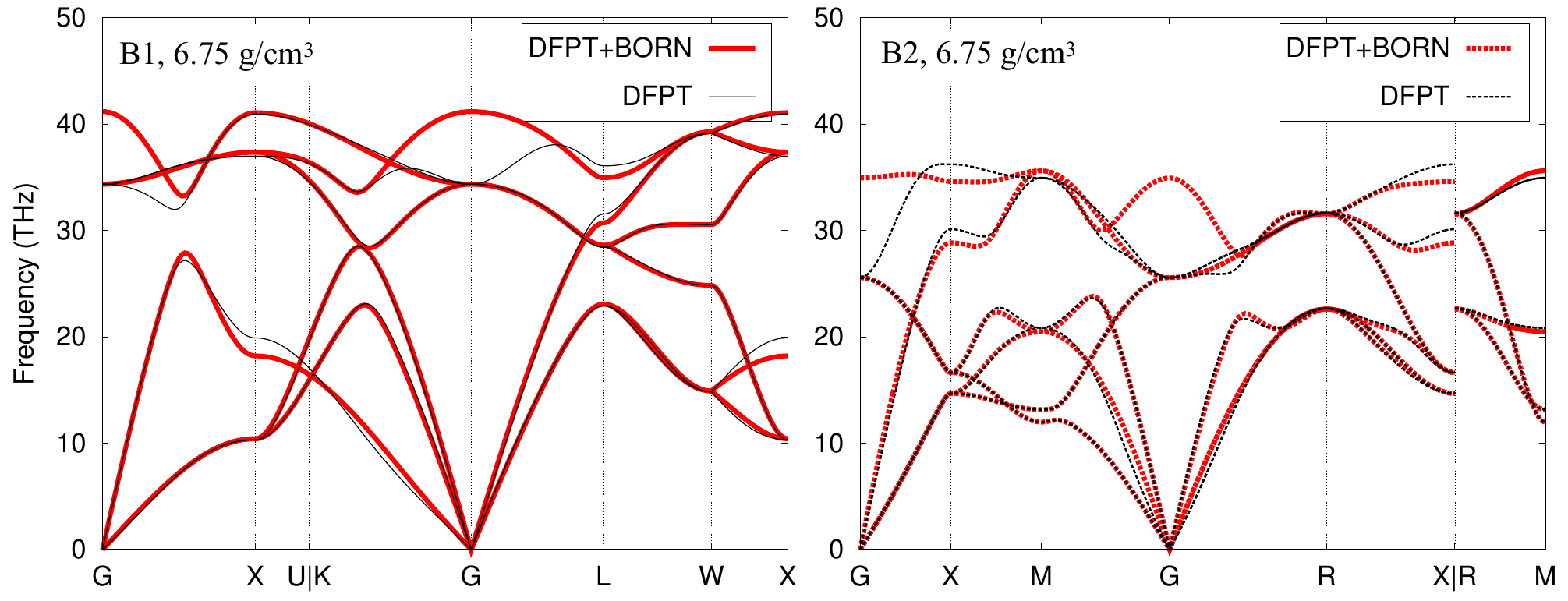}
\caption{Phonon band structures of MgO from DFPT vs DFPT+Born calculations of B1 vs B2 MgO at 6.75~g/cm$^3$.}
\label{fig:lotophonon}
\end{figure*}

It is well known that the frequencies of the optical modes parallel and perpendicular to the electric field split (``LO-TO splitting'') in ionic materials such as MgO.~\cite{Alfe2009}
This mode splitting is missed in regular phonon calculations but can be correctly captured when Born effective charges, piezoelectric constants, and the ionic contribution to the dielectric tensor are considered (by switching on {\footnotesize LEPSILON} in {\footnotesize VASP}). 
The effects on the phonon dispersion relations of B1- and B2-MgO are shown in Fig.~\ref{fig:lotophonon}.

\begin{figure*}[ht]
\centering
\includegraphics[width=0.9\linewidth]{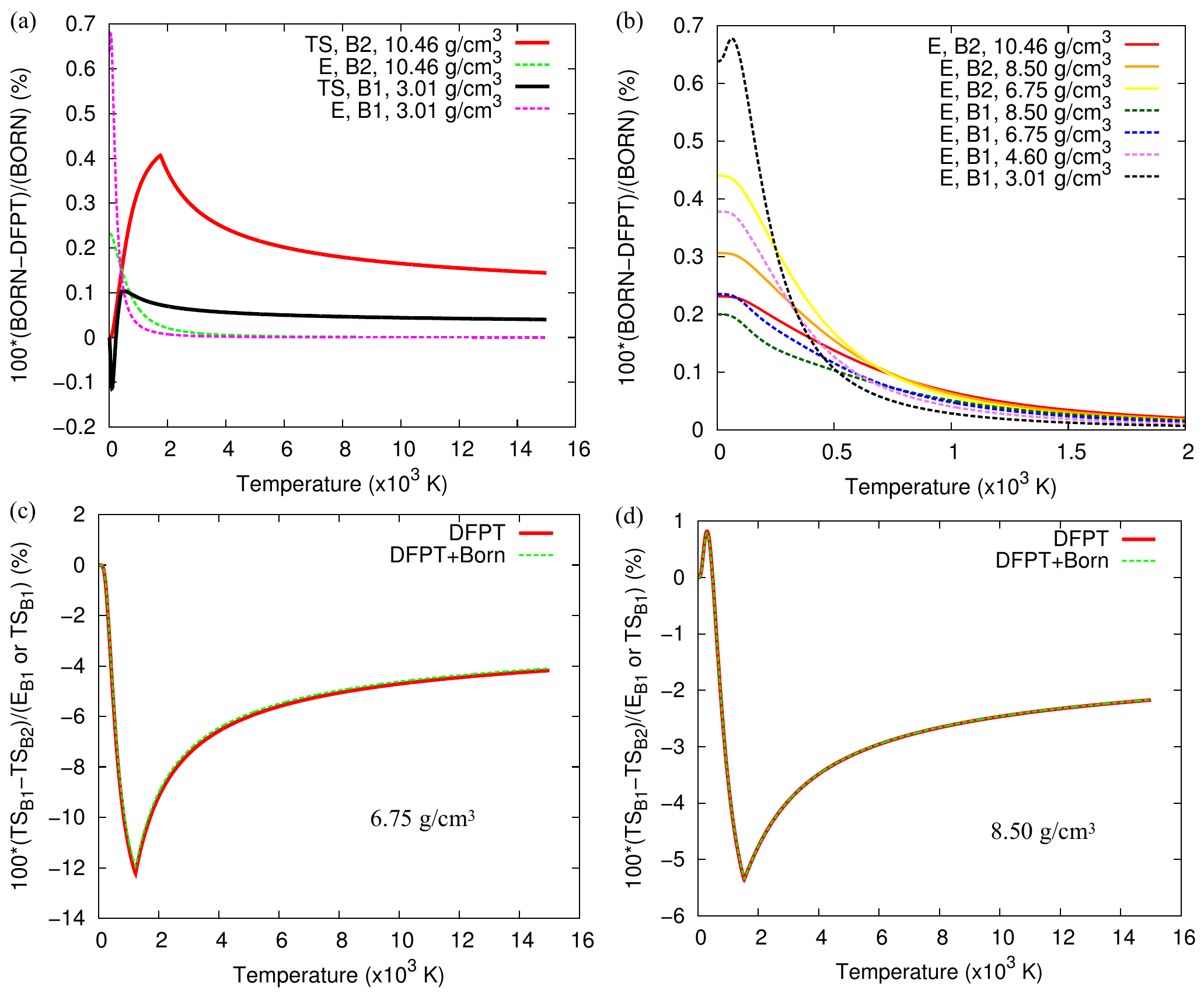}
\caption{The effects of including LO-TO splitting (BORN) in phonon calculations on the thermodynamic properties of B1 and B2 phases of MgO at various densities. (a) Entropy and internal energy and (b) internal energy changes due to excluding LO-TO splitting; (c) and (d) entropy differences between B1 and B2 at two densities near the phase transition with and without LO-TO splitting. The differences in (c) and (d) are relative to $E$ or $TS$, whichever is larger.}
\label{fig:lotophononETS}
\end{figure*}

Figure~\ref{fig:lotophononETS} compares the resultant differences in vibrational energy and entropy of MgO in different phases and at different densities. The results show that LO-TO splitting only makes a small difference ($<$0.7\%) at $T<500$ K and then quickly drops to zero at higher T; 
the effect on the differences between B1 and B2 is also small and negligible.

\section{Effects of xc functional on phonon results}\label{secapp:xc}

\begin{figure*}[ht]
\centering
\includegraphics[width=0.8\linewidth]{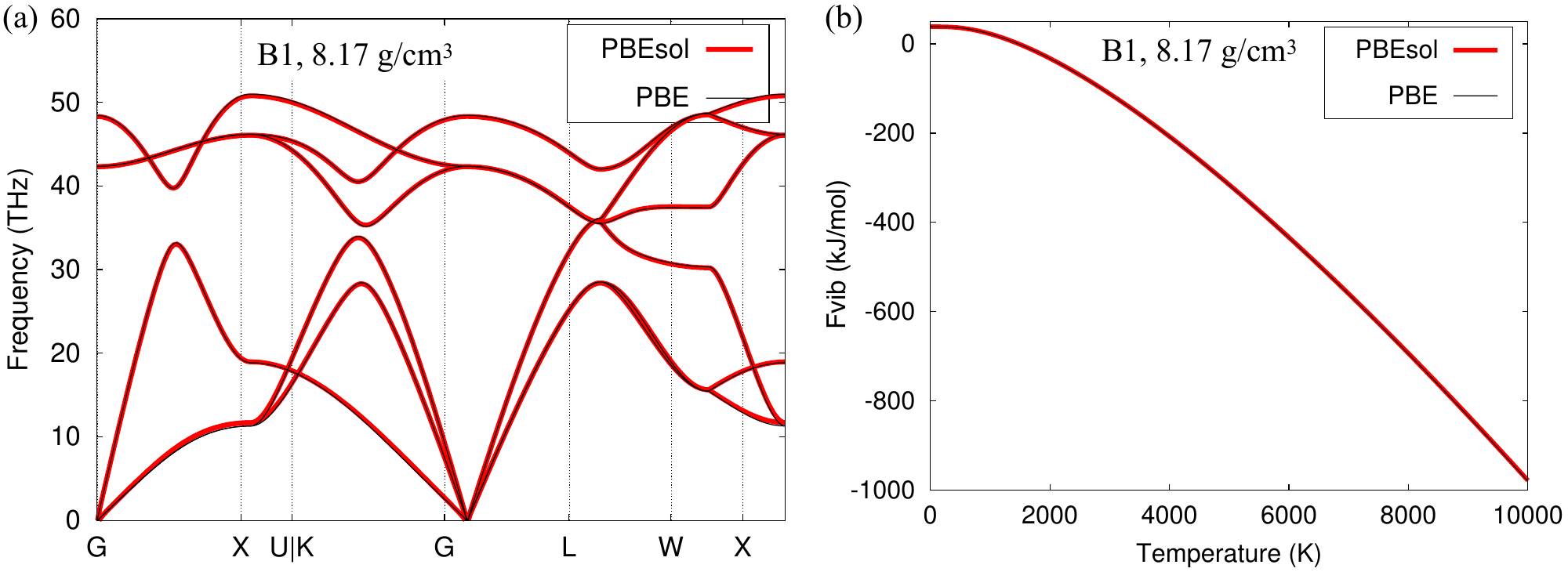}
\caption{(a) Phonon band structures and (b) vibrational free energies of MgO from PBEsol vs PBE calculations.}
\label{fig:xcphonon}
\end{figure*}

Figure~\ref{fig:xcphonon} shows that different xc functionals produce the same phonon band structure and vibrational free energies within QHA.

\section{Convergence test}\label{secapp:convergence}

\begin{figure*}[ht]
\centering
\includegraphics[width=0.9\linewidth]{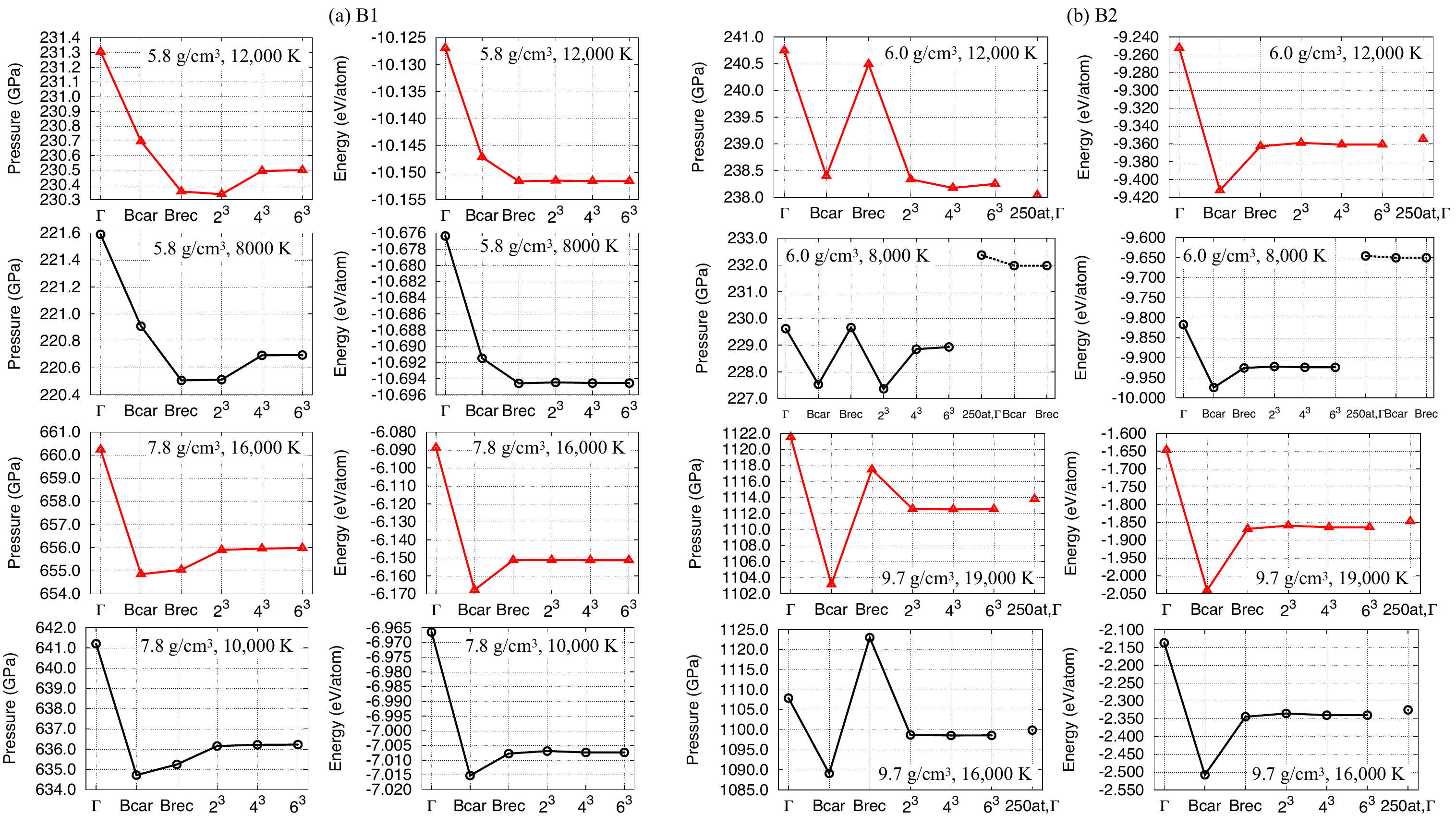}
\caption{Finite size effects on pressures and internal energies of (a) B1 and (b) B2 structures of MgO at different densities and electronic temperatures. 
All calculations are based on DFT-PBEsol and use 64-atom cells for B1 and 54-atom cells for B2, unless otherwise specified. ``Bcar'' and ``Brec'' denotes using the special $k$ point of $(1/4,1/4,1/4)$ in cartesian (wrong ``Baldereschi point'' for cubic cell) and reciprocal (correct ``Baldereschi point'' for cubic cell) coordinates, respectively.}
\label{fig:converge}
\end{figure*}

Figure~\ref{fig:converge} shows large cell sizes in combination with proper/fine $k$-point meshes are needed to ensure convergence of the EOS. For example, a 250-atom cell with a single $k$ point is not enough for B2 at 6.0~g/cm$^3$.
In our QMD simulations, we use a 64-atom cell with the ``Brec'' special $k$ point and a 54-atom cell with a $\Gamma$-centered 2$\times$2$\times$2 $k$ mesh, respectively, for the B1 and B2 calculations.

Our additional tests for the phonon calculations show that 8- and 16-atom cells, respectively, are needed for B1 and B2 to obtain converged $F_\text{vib}(T)$ results rather than using the primitive 2-atom cells. 
In this study, we choose 54-atom cells (with a 4$\times$4$\times$4 $k$ mesh) for both B1 and B2 phonon calculations for better accuracy.

\section{Calculation of anharmonic energies and comparison between different terms}\label{secapp:Fah}

\begin{figure*}[ht]
\centering
\includegraphics[width=0.95\linewidth]{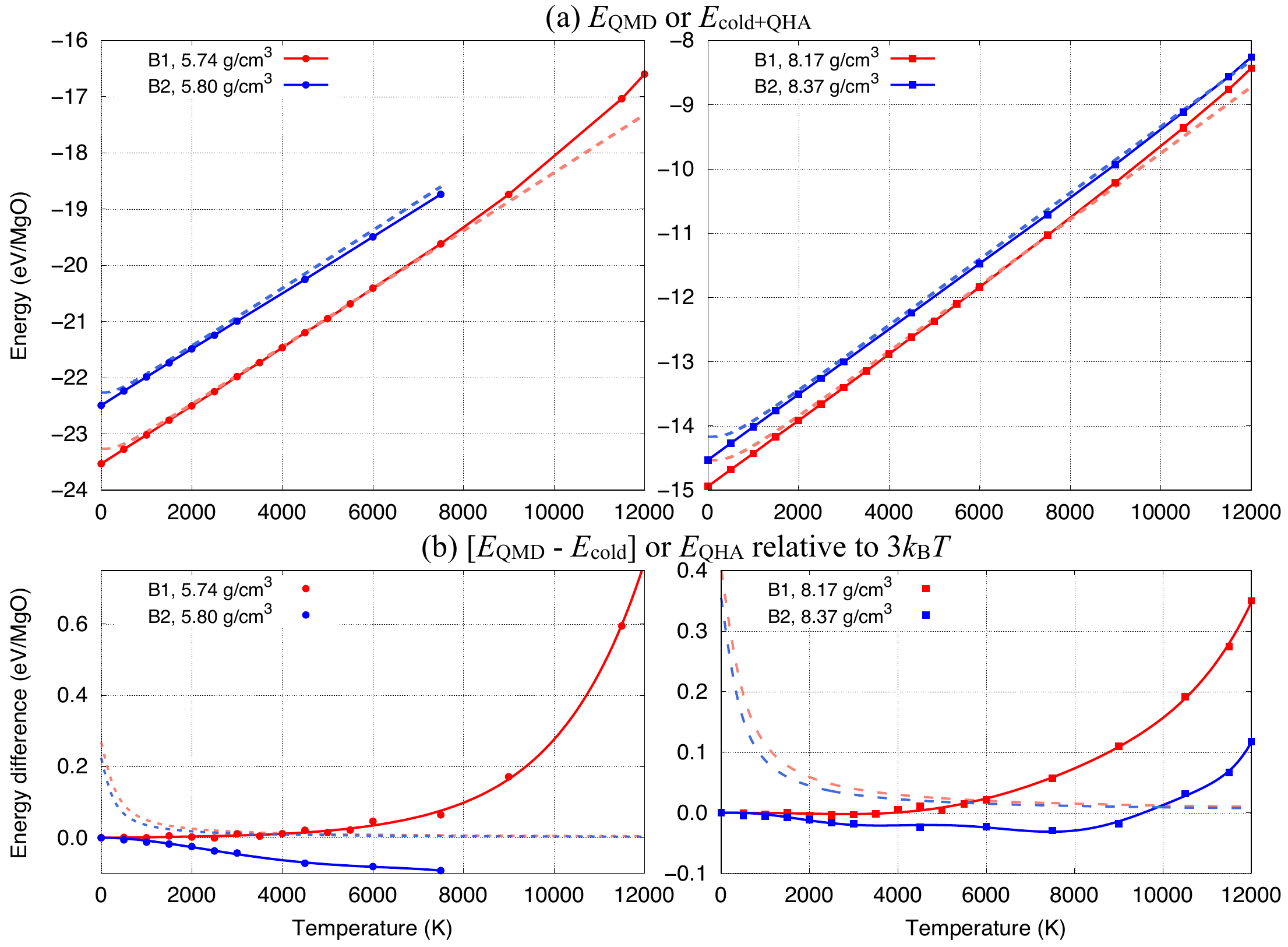}
\caption{(a) Internal energies along selected isochores of MgO based on QMD (darker solid curves) or QHA (lighter dashed curves) calculations. (b) Differences of the ion thermal term of the internal energy (QMD: solid curves and symbols; QHA: dashed curves) from a classical crystal that assumes 3$k_\text{B}$/atom for the heat capacity (the Dulong--Petit law). The structures and densities are represented by different colors and symbols as denoted in the legend. The solid curves in (b) are polynomial fits to the data. In both panels, $E_\text{cold}$ is taken as the value of $E_\text{QMD}$ at 0 K.}
\label{fig:supcalcFah}
\end{figure*}

\begin{figure*}[ht]
\centering
\includegraphics[width=0.95\linewidth]{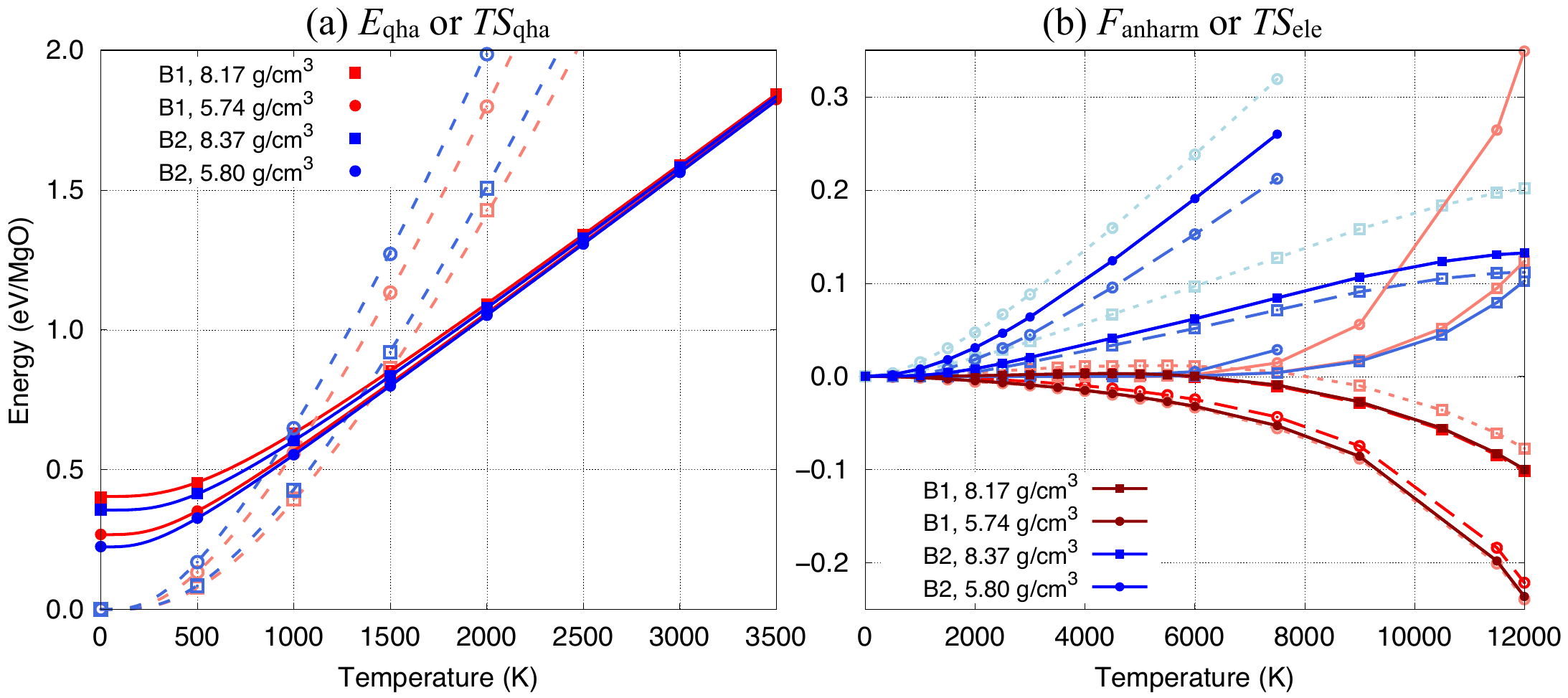}
\caption{Comparison between different thermodynamic energy terms of MgO, including (a) internal energy (solid curves) and vibration entropy (dashed curves) terms under QHA calculations, and (b) anharmonic free energy (dark solid and light dashed curves) and electronic entropy (light solid curves) terms from TDI and QMD calculations. The anharmonic free energies shown in (b) are calculated using TDI with different methods: polynomial fit of anharmonic internal energy ($E_\text{anharm}=E_\text{QMD}-E_\text{cold}-3k_BT$) with $C_V(T_\text{ref})$=10\%$\times3k_\text{B}$/atom (dark solid{\color{black}, corresponding $T_\mathrm{ref}$=100--200~K}) or 90\%$\times3k_\text{B}$/atom (light long dashed{\color{black}, corresponding $T_\mathrm{ref}$=800--1550~K}) or cubic spline fit of the internal energy ($E_\text{QMD}$) with $C_V(T_\text{ref})$=50\%$\times3k_\text{B}$/atom (light short dashed{\color{black}, corresponding $T_\mathrm{ref}$=250--550~K}).}
\label{fig:supFterms}
\end{figure*}

Figure~\ref{fig:supcalcFah} shows the finite-temperature internal energies of MgO estimated from cold calculations under QHA ($E_\text{cold+QHA}$) in comparison with values from direct QMD simulations ($E_\text{QMD}$).
Overall, $E_\text{cold+QHA}$ and $E_\text{QMD}$ are similar to each other, with noticeable differences near zero K, because of the nuclear quantum effects, or at high temperatures, due to increased anharmonic vibration and electron excitation effects. 
The differences are more evident when the ion thermal energies ($E-E_\text{cold}$) are plotted with respect to the classical crystal value of $3k_BT$.
The mismatch between QMD and QHA near zero K and the proximity of $E_\text{QHA}$ to $3k_BT$ at high temperatures have motivated us to define $E_\text{anharm}=E_\text{QMD}-E_\text{cold+QHA}\approx E_\text{QMD}-E_\text{cold}-3k_BT$ in the TDI Eq.~\ref{eq:tdi} to calculate the anharmonic free energy $F_\text{anharm}$.
Under this approximation, the total free energy of the system 
$F(V, T)=E_{\mathrm{cold}}(V)+F_{\mathrm{i}, \mathrm{QHA}}(V, T)+F_{\mathrm{i}-\mathrm{th}, \mathrm{anharm}}(V, T) +F_{\mathrm{e}-\mathrm{{\color{black}th}}}(V, T)
\approx E_{\mathrm{cold}}(V)+F_{\mathrm{ind.ph.}}^{\mathrm{quantum.}}(V, T)+F_{\mathrm{int.ph.}}^{\mathrm{class.}}(V, T) +F_{\mathrm{e}-\mathrm{{\color{black}th}}}(V, T)$
where the subindices ``ind.'' and ``int.'' denote independent and interacting, ``ph.'' denotes phonon, ``class.'' and ``quantum.'' represents the nature of the ions as being classical and quantum, respectively, {\color{black}and ``e-th'' denotes the electron thermal term}.
The only difference from an entirely accurate (``quantum'') description lies in the approximation in the anharmonic term by using classical ions (as in QMD simulations and the classical-crystal reference for TDI), whose effect, we believe, is negligible for the purpose of this paper.
We have performed extensive tests and found $E_\text{anharm}(V,T)$ can be isochorically fitted well using sixth- and eighth-order polynomials for the B1 and B2 phases, respectively.
We also note that the different choices of $T_\text{ref}$ or fitting $E_\text{QMD}$ by using cubic splines can affect the value of $F_\text{anharm}$ (see Fig.~\ref{fig:supFterms}), while lower-order polynomials or exponential fits~\cite{Oganov2004,Sun2014exp}, although they were found to work for certain materials at ambient densities or relatively low temperatures, break down for MgO at high densities and temperatures.

In practice, QMD is less efficient and inappropriate for simulating near-zero temperatures. Therefore we have to choose a finite value for $T_\text{ref}$ in TDI and assume QHA is valid for any temperature below $T_\text{ref}$. This would technically limit the accuracy of the anharmonic free energies, as shown in Fig.~\ref{fig:supFterms}(b) by the different values of $F_\text{anharm}$ when choosing different $T_\text{ref}$ and fitting approaches. 
Figure~\ref{fig:supFterms} also shows that the contributions by  electron thermal excitation become increasingly significant when the temperature exceeds 8000 K, 
more at lower densities.
The anharmonic vibration and electron thermal terms are relatively small in comparison to the lattice vibration as accounted under QHA. 
However, because of the similarities between energies of the B1 and B2 phases, the effects of anharmonic vibration can significantly affect the B1-B2 transition boundary, as shown in Fig.~\ref{fig:B1B2trans}.

\begin{figure}[ht]
\centering
\includegraphics[width=0.8\linewidth]{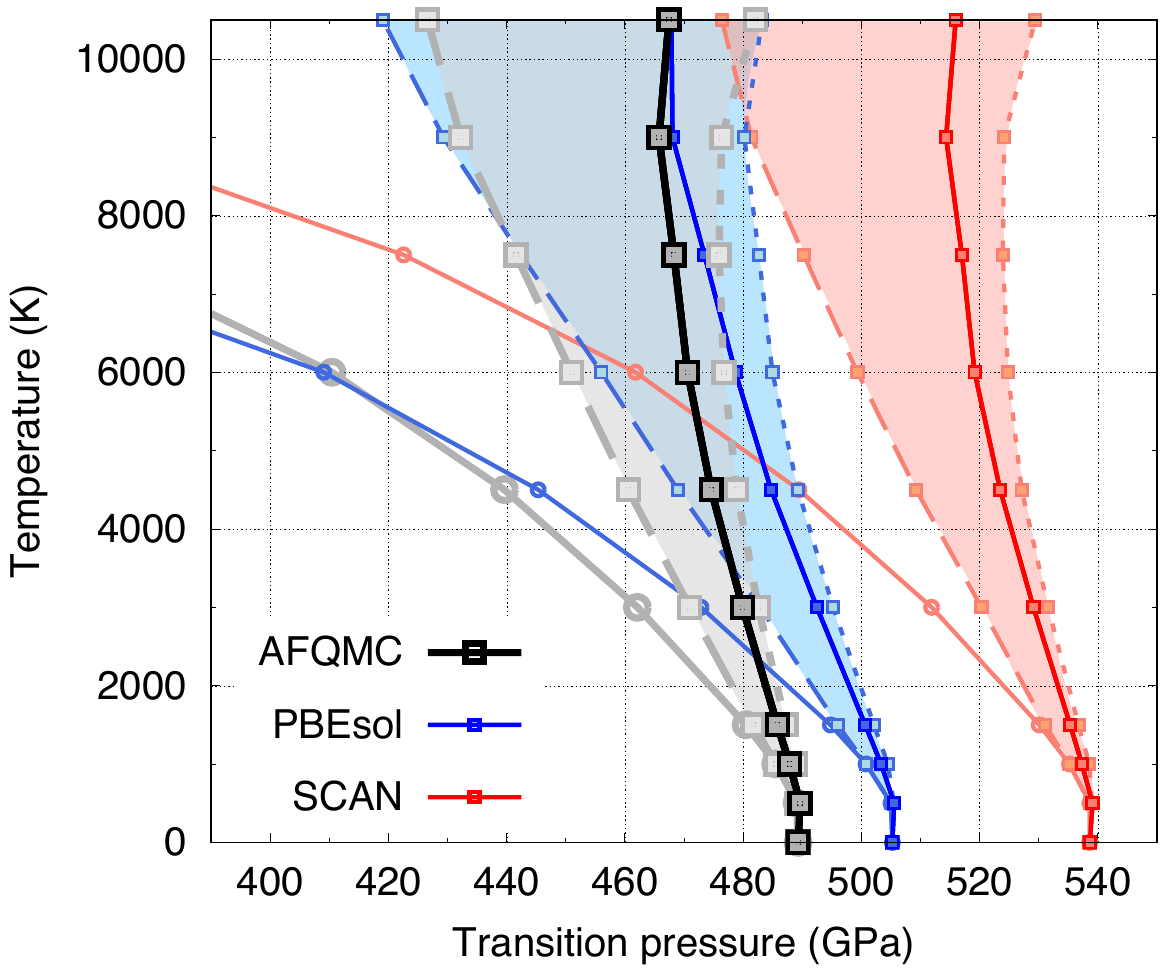}
\caption{B1-B2 phase boundary calculated using QHA (light solid line-circles) in comparison to those including the anharmonic effect calculated with three different methods in TDI: polynomial fit of $E_\text{anharm}$ with $C_V(T_\text{ref})$=10\%$\times3k_\text{B}$/atom (dark solid line-squares{\color{black}, corresponding $T_\mathrm{ref}$=100--200~K}) or 90\%$\times3k_\text{B}$/atom (light long dashed line-squares{\color{black}, corresponding $T_\mathrm{ref}$=800--1550~K}) or cubic spline fit of the internal energy ($E_\text{QMD}$) with $C_V(T_\text{ref})$=50\%$\times3k_\text{B}$/atom (light short dashed line-squares{\color{black}, corresponding $T_\mathrm{ref}$=250--550~K}). The maximum range of difference defined by the three methods are represented by the shaded area.
Black, blue, and red colors denote the calculations that use different cold curves.}
\label{fig:supPtr}
\end{figure}

Figure~\ref{fig:supPtr} summarizes the B1-B2 transition pressure based on free energies calculated using different approaches. 
Despite the distinctions between predictions by AFQMC and DFT-PBEsol or SCAN at zero K, all methods give similar trends of decreasing $P_\text{tr}$ (by $\sim$20 to 40 GPa at 9000 K, relative to the corresponding values at 0 K) and enlarging uncertainty (by $\sim$40 to 50 GPa at 9000 K) as temperature increases.
The relations in $P_\text{tr}$ between the different approaches AFQMC, DFT-PBEsol, and DFT-SCAN at high temperatures remain similar to those under QHA, whereas the anharmonic effects clearly steepen the $\mathrm{d}T/\mathrm{d}P$ slope and push $P_\text{tr}$  to higher values than QHA predictions.
With polynomial fits of $E_\text{anharm}$, the differences between the phase boundaries based on QHA and anharmonic calculations are smaller if the values of $T_\text{ref}$ are higher (light long dashed line-squares); in comparison, cubic spline fits of $E_\text{QMD}$ using the same $T_\text{ref}$ tend to produce larger differences than the polynomial fits of $E_\text{anharm}$ (light short dashed line-squares).
{\color{black}We have quantified the phase-boundary differences by calculating the Clapeyron slope. The results are summarized in Table~\ref{tab:dPdTtr}.}

\begin{table}[h]
{\color{black}
\caption{\label{tab:dPdTtr} Clapeyron slope $dP_\mathrm{tr}/dT$ (in units of MPa/K) of the MgO B1--B2 phase transition estimated by linear regression of the data calculated using different approaches as shown in Fig.~\ref{fig:supPtr}. The calculations for QHA use 1500--4500 K data, while other cases use 1500--6000 K data, for better linearity and relevance to the super-Earths' interior conditions. Numbers in boldface correspond to the solid line squares in Figs.~\ref{fig:B1B2trans}(b) and~\ref{fig:supPtr}. We note that the values of the slope calculated here are much smaller than previous estimations based on experiments ($-390\pm300$~MPa/K)~\cite{McWilliamsScience2012}, which were associated with significant uncertainties, and also lower than predictions by an interatomic model of the B1-B2 transition ($\sim-40$~MPa/K if taking volume collapse of 3\% and entropy increase of 7~J/K/mol)~\cite{B1B2modelJeanloz1982,Jeanloz1982}, suggesting the B1-B2 entropy difference and the thermodynamic properties of MgO are sensitive to pressure and different from the underlying assumptions of the model.}
    \centering
    \scriptsize
    \begin{ruledtabular}
    \begin{tabular}{rcccc}
         & QHA & $C_V(T_\text{ref})$=10\%$\times3k_\text{B}$/atom & $C_V(T_\text{ref})$=90\%$\times3k_\text{B}$/atom & $C_V(T_\text{ref})$=50\%$\times3k_\text{B}$/atom \\
         \hline
AFQMC & -13.6 $\pm$ 0.8 & {\bf -3.4 $\pm$ 0.2} & -6.8 $\pm$ 0.1 & -2.3 $\pm$ 0.3 \\
SCAN & -13.5 $\pm$ 0.8 & {\bf -3.6 $\pm$ 0.2} & -7.1 $\pm$ 0.1 & -2.7 $\pm$ 0.3 \\
PBEsol & -16.5 $\pm$ 1.1 & {\bf -4.9 $\pm$ 0.2} & -8.9 $\pm$ 0.1 & -3.8 $\pm$ 0.3 \\
    \end{tabular}
    \end{ruledtabular}
}
\end{table}

{\color{black}Furthermore, we have tested by employing two different versions of the TDI/temperature-integration approach to cross-check our above results, including
(1) a more direct approach~\cite{VanGunsteren2002}:
\begin{equation} 
\frac{F(V, T)}{T}-\frac{F(V, T_{\mathrm{ref}})}{T_{\mathrm{ref}}}= \int_{1/T_{\mathrm{ref}}}^{1/T} E(V, \mathcal{T}) 
\mathrm{d} \frac{1}{\mathcal{T}}\label{eq:tdi2direct}
\end{equation}
and (2) an indirect approach [by taking the difference of Eq.~\ref{eq:tdi2direct} with respect to a reference system (e.g., the system under QHA) that also satisfies Eq.~\ref{eq:tdi2direct}]:
\begin{equation} 
F(V, T)-F_{\mathrm{ref}}(V, T)= T\int_{1/T_{\mathrm{ref}}}^{1/T} [E(V, \mathcal{T})-E_{\mathrm{ref}}(V, \mathcal{T}) ]
\mathrm{d} \frac{1}{\mathcal{T}}.\label{eq:tdi2indirect}
\end{equation}

These tests were performed at $T=$3000, 6000, and 9000~K, with $T_\mathrm{ref}$ fixed to 500~K for simplicity.
In approach (1), the free energy at $T_{\mathrm{ref}}$ is approximated by the corresponding values under QHA; in approach (2), the QHA system is taken as the reference, which defines $F_{\mathrm{ref}}$ and $E_{\mathrm{ref}}$.
In both approaches, an additional term $E_{\mathrm{QC}}(V, \mathcal{T})=E_{\mathrm{QHA}}(V, \mathcal{T})-3k_{\mathrm{B}}\mathcal{T}$ has been introduced as quantum correction of the internal energy from QMD, similar to that in Ref.~\onlinecite{Jung2023}. We note that the quantum correction is crucial to obtain accurate free energies within the temperature integration approach, which starts from a cold reference state where the important nuclear quantum effects are included by QHA but missed in QMD. We also note that, with the quantum correction and with $E_{\mathrm{cold}}$ deducted from all energy terms, Eq.~\ref{eq:tdi2indirect} is equivalent to our method introduced in detail above and in Sec.~\ref{subsec2:highT} (Eq.~\ref{eq:tdi}).

Based on our PBEsol data (cold curve, QHA, and QMD), the free energies calculated using these two approaches are similar, both producing similar B1--B2 transition pressures: 486 GPa at 3000 K, 462 GPa at 6000 K, and 439 GPa at 9000 K. The excellent consistency of the results from the tests with those shown in Fig.~\ref{fig:supPtr} (blue shaded area) reconfirms the methodology and findings of this study.
}

\begin{figure}[ht]
\centering
{\color{black}
\includegraphics[width=0.8\linewidth]{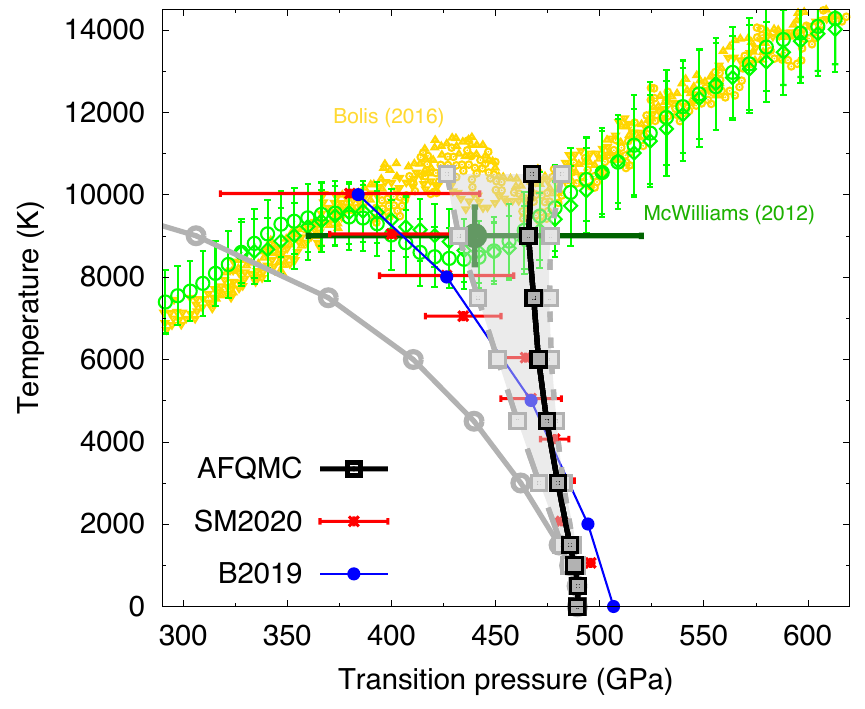}
\caption{The B1--B2 phase boundary of MgO calculated in this study (same as the black/grey AFQMC results shown in Fig.~\ref{fig:supPtr}) in comparison to shock experiments by McWilliams {\it et al.}~\cite{McWilliamsScience2012} (green symbols) and Bolis {\it et al.}~\cite{Bolis2016mgo,noteaboutBolis2016} (yellow symbols) and recent theoretical calculations by Bouchet {\it et al.}~\cite{Bouchet2019} (temperature-dependent effective potential approach, LDA xc functional, data represented by blue line-circles) and Soubiran and Militzer~\cite{Soubiran2020} (TDI based on MD using effective potentials tuned between harmonic oscillators and Kohn--Sham DFT, PBE xc functional, data shown with red crosses). The dark-green circle denotes the condition attributed to the B1--B2 transition by McWilliams {\it et al.}~\cite{McWilliamsScience2012}.}
\label{fig:supPtrvsExpt}
}
\end{figure}

{\color{black}
\section{Comparison with experiments}\label{secapp:PtrvsExpt}
}

{\color{black}
Figure~\ref{fig:supPtrvsExpt} compares our AFQMC results of the B1--B2 transition to shock experiments~\cite{McWilliamsScience2012,Bolis2016mgo} and recent theoretical calculations~\cite{Bouchet2019,Soubiran2020}.
The previous calculations were based on LDA/PBE, and their predicted $P_{\mathrm{tr}}$ at 0 K is larger than our AFQMC prediction, consistent with our findings shown in Fig.~\ref{fig:trans0K}.
The differences get smaller at higher temperatures until approximately 8000~K, above which the previous calculations show a lower $P_{\mathrm{tr}}$ and thus a less steep Clapeyron slope than ours.
Our estimation of the B1--B2 phase boundary, with uncertainty, agrees with the wiggled regions in both experiments by McWilliams {\it et al.}~\cite{McWilliamsScience2012} and Bolis {\it et al.}~\cite{Bolis2016mgo}.
This suggests the turnovers in both experiments can be associated with the B1--B2 transition.
To fully unveil the origins of the subtle differences between the measurements, however, still requires improved experimental diagnostics and theoretical constraints on the structure, kinetics, and thermodynamic conditions of the samples under shock compression.
}


\begin{thebibliography}{95}
\bibitem [{not()}]{noteaboutBolis2016}%
  \BibitemOpen
  \href@noop {} {}\bibinfo {note} {{\color{black}Note that the experimental
  data by Bolis {\it et al.} shown here are scanned from Fig.~8 of
  Ref.~\cite{Bouchet2019} and lifted by 1300 K to approximately match the
  original publication (Fig. 2 of Ref.~\cite{Bolis2016mgo}). Unfortunately, the
  original data in Ref.~\cite{Bolis2016mgo} are not available and it is unclear
  where the mismatch was from.}}\BibitemShut {Stop}%
\end{thebibliography}

%

\end{document}